\documentclass[aps,prl,superscriptaddress,twocolumn,citeautoscript, longbibliography]{revtex4-1}
\usepackage{graphicx}
\usepackage{amsfonts}
\usepackage{amsmath}
\usepackage{amssymb}
\usepackage{bm}
\usepackage{textcomp}
\usepackage{verbatim}
\usepackage{xspace}
\usepackage{soul}
\usepackage{bbold}
\usepackage{mathtools}
\usepackage{dcolumn}
\usepackage{hyperref}
\usepackage[usenames]{color}
\usepackage{subfig}
\usepackage{physics}
\usepackage{units}
\usepackage[export]{adjustbox}
\usepackage{tablefootnote}
\usepackage[dvipsnames]{xcolor}
\usepackage{ragged2e}
\DeclareCaptionJustification{justified}{\justifying}
\def\beq{\begin{equation}}
\def\eeq{\end{equation}}
\def\vk{{\bf k}}

\def\vR{{\bf R}}

\def\jeff{{$J_{\text{eff}}$ \xspace}}

\newcommand{\out}[1]{{}}

\newcommand{\xs}{$x^2$-$y^2$\xspace}
\newcommand{\zs}{$z^2$\xspace}

\newcommand{\bmro}{Ba$_2$MgReO$_6$ }

\usepackage{graphicx}
\usepackage{amsmath}
\usepackage{tikz}
\usepackage{subcaption}
\usetikzlibrary{arrows.meta,calc}

\definecolor{ligandfill}{RGB}{255,255,255}   
\definecolor{ligandline}{RGB}{ 40, 40, 40}   
\definecolor{octfacecol}{RGB}{170,170,170}   
\definecolor{octedgecol}{RGB}{130,130,130}   
\definecolor{modeU}{RGB}{ 45,110,230}        
\definecolor{modeV}{RGB}{255,127,0}        
 
\tikzset{
  octaxes/.style={x={(0.940cm,-0.089cm)}, y={(0.342cm,0.243cm)}, z={(0cm,0.966cm)}},
  octface/.style={fill=octfacecol, fill opacity=0.13, draw=none},
  ligand/.style={circle, draw=ligandline, fill=ligandfill, inner sep=0pt},
}
 
\newcommand{\octbbxfac}{1.80}
\newcommand{\octbbyfac}{1.98}
 
\newcommand{\octahedron}[1][\jtscale]{%
  \pgfmathsetmacro{\octbbx}{\octbbxfac*#1}%
  \pgfmathsetmacro{\octbby}{\octbbyfac*#1}%
  \pgfmathsetmacro{\octligd}{1.7*#1}
  \pgfmathsetmacro{\octligw}{0.58*#1}
  \pgfmathsetmacro{\octedgw}{0.32*#1}
  \path[use as bounding box]
      (-\octbbx cm,-\octbby cm) rectangle (\octbbx cm,\octbby cm);
  \begin{scope}[octaxes, scale=#1,
                ligand/.append style={minimum size=\octligd mm,
                                      line width=\octligw pt}]
    \coordinate (Lxp) at ( 1, 0, 0);   \coordinate (Lxm) at (-1, 0, 0);
    \coordinate (Lyp) at ( 0, 1, 0);   \coordinate (Lym) at ( 0,-1, 0);
    \coordinate (Lzp) at ( 0, 0, 1);   \coordinate (Lzm) at ( 0, 0,-1);
    \foreach \a/\b in {Lxp/Lyp, Lxm/Lyp, Lxp/Lym, Lxm/Lym}{
      \fill[octface] (\a) -- (\b) -- (Lzp) -- cycle;
      \fill[octface] (\a) -- (\b) -- (Lzm) -- cycle;}
    \begin{scope}[draw=octedgecol, line width=\octedgw pt]
      \begin{scope}[draw opacity=0.35]
        \draw (Lxp) -- (Lyp) -- (Lxm);
        \draw (Lyp) -- (Lzp);  \draw (Lyp) -- (Lzm);
      \end{scope}
      \foreach \a in {Lxp,Lxm,Lym}{\draw (\a) -- (Lzp);  \draw (\a) -- (Lzm);}
      \draw (Lxp) -- (Lym) -- (Lxm);
    \end{scope}
    \node[ligand, draw opacity=0.30] at (Lyp) {};
    \foreach \v in {Lxp,Lxm,Lym,Lzp,Lzm}{\node[ligand] at (\v) {};}
  \end{scope}}
 
\newcommand{\arrowalong}[5][\jtscale]{%
  \pgfmathsetmacro{\octahl}{1.94*#1}
  \pgfmathsetmacro{\octahw}{1.68*#1}
  \pgfmathsetmacro{\octalw}{1.16*#1}
  \begin{scope}[octaxes, scale=#1]
    \draw[-{Stealth[length=\octahl mm,width=\octahw mm]},
          line width=\octalw pt, color=#2] ($#4*(#3)$) -- ($#5*(#3)$);
  \end{scope}}
 
\newcommand{\octaEu}[1][\jtscale]{%
  \begin{tikzpicture}
    \octahedron[#1]
    \arrowalong[#1]{modeU}{0,0, 1}{1.18}{1.80}
    \arrowalong[#1]{modeU}{0,0,-1}{1.18}{1.80}
    \arrowalong[#1]{modeU}{ 1,0,0}{1.66}{1.19}
    \arrowalong[#1]{modeU}{-1,0,0}{1.66}{1.19}
    \arrowalong[#1]{modeU}{0, 1,0}{1.66}{1.19}
    \arrowalong[#1]{modeU}{0,-1,0}{1.66}{1.19}
  \end{tikzpicture}}
 
\newcommand{\octaEv}[1][\jtscale]{%
  \begin{tikzpicture}
    \octahedron[#1]
    \arrowalong[#1]{modeV}{ 1,0,0}{1.18}{1.85}
    \arrowalong[#1]{modeV}{-1,0,0}{1.18}{1.85}
    \arrowalong[#1]{modeV}{0, 1,0}{1.85}{1.18}
    \arrowalong[#1]{modeV}{0,-1,0}{1.85}{1.18}
  \end{tikzpicture}}

\begin{document}

\title{Origin of quadrupolar order in the 5d$^1$ double perovskite \bmro}

\author{Dario Fiore Mosca}
\address{University of Vienna, Faculty of Physics, Center for Computational Materials Science, Kolingasse 14-16, 1090, Vienna, Austria}

\author{Leonid V. Pourovskii}
\address{CPHT, CNRS, \'Ecole polytechnique, Institut Polytechnique de Paris, 91120 Palaiseau, France}
\address{Coll\`ege de France, Université PSL, 11 place Marcelin Berthelot, 75005 Paris, France}

\begin{abstract}
Despite extensive experimental and theoretical studies of quadrupolar order in Ba$_2$MgReO$_6$, the microscopic origin and relative energy scales of the interactions driving it remain unresolved. In particular, the observed ferroic ordering of the \zs quadrupole has remained difficult to explain, as the relative contributions of superexchange, classical quadrupole-quadrupole interactions, vibronic couplings, and virtual excitations to higher-energy levels have not been established quantitatively. Here, we determine all 
these interactions from first principles and solve the resulting microscopic Hamiltonian. The calculated ground state reproduces the experimentally observed quadrupolar order parameters and the associated structural distortions with quantitative accuracy. Disentangling the individual contributions, we find that ferroic \zs order originates from a linear-quartic coupling, generated by virtual excitations to the \jeff= 1/2 doublet and driven by vibronic interactions.
Our results establish the hierarchy of microscopic interactions underlying quadrupolar order in \bmro and provide a unified \emph{ab initio} description of its coupled electronic and lattice degrees of freedom.
\end{abstract}

\maketitle

{\it Introduction. --- } Interactions between the microscopic degrees of freedom in condensed matter determine the phases and phenomena that emerge at macroscopic scales. Particularly rich behavior arises when several interactions compete or cooperate, giving rise to phenomena such as quantum spin liquids~\cite{Balents2010}, unconventional superconductivity~\cite{Scalapino2012}, multiferroicity~\cite{Spaldin2019} and multipolar orders~\cite{Santini2009, Kuramoto2009}. Understanding the hierarchy and interplay of low-energy interactions is therefore central for identifying the microscopic origins of emergent quantum phases.

Double perovskites containing heavy transition-metal ions provide an ideal setting in which to explore this interplay. In these materials, strong single-ion interactions like Coulomb's $U$, crystal field (CF),  Hund's exchange $J_H$ and spin-orbit coupling (SOC)  select a low-lying ground state multiplet (GSM)  based on the electronic configuration~\cite{Takayama2021, Pourovskii2025}. Within this GSM, weaker intersite interactions compete to determine the collective ground state. These include hopping-mediated intersite exchange interactions  (IEI)  like direct exchange or superexchange, electrostatic or classical quadrupole-quadrupole (cQQ) repulsion~\cite{Chen2010} and vibronic interactions in the form of intersite elastic couplings (IEC) and Jahn-Teller effect (JT)~\cite{Iwahara2024}. Because these energy scales can be comparable, the resulting order cannot, in general, be attributed to a single microscopic mechanism~\cite{Witczak2014}. 

The relative importance of these interactions is central for understanding the $5d^1$ double perovskite \bmro (BMRO). Here, the CF effect promotes the occupation of the $t_{2g}$ orbitals, well separated in energy from the $e_g$ ones ($\sim$ 4.5 eV), and that carry an effective orbital angular momentum $\tilde{l} = 1$. Strong SOC then splits this manifold into a \jeff = 3/2 GSM and an excited \jeff$=1/2$ multiplet separated by $3\lambda/2$, where $\lambda$ is the SOC strength ($\sim $ 0.3 eV). Owing to its spin-orbital character, the \jeff$=3/2$ GSM can host multipolar moments beyond conventional dipoles, up to rank $K = 2$\jeff. Moreover, being formed by two Kramers doublets, it  is JT active~\cite{Iwahara2018}. The low-energy degrees of freedom thus comprise both electronic and magnetic multipoles as well as lattice distortions, whose interactions can stabilize a variety of ordered states.

In BMRO this is evidenced by the formation of a composite quadrupolar ordered phase 
initially identified from structural refinement by Hirai and coworkers~\cite{Hirai2019_JPSJ, Hirai2020_PRR}. 
This order is characterized by the simultaneous onset of antiferroic quadrupole $x^2-y^2$ (AF$x^2$-$y^2$) with wavevector $\vk = [0\ 0\ 1]$ and ferroic \zs  (F$z^2$)~\cite{Hirai2020_PRR}. More recently, Soh and coworkers observed clear signatures of both order parameters in Resonant Elastic X-ray Scattering~\cite{Soh2024}, establishing the simultaneous ordering of the electronic quadrupoles and the associated structural distortions.
The quadrupole order onset is signaled by a broad specific-heat anomaly at  $T_q \approx 33$ K~\cite{Marjerrison2016}, but the expected  $\ln(4)$ entropy of the \jeff quadruplet is not recovered well above $T_q$~\cite{Pastrova2023}. This  suggests that above $T_q$ BMRO 
might form a glassy phase, recently proposed for the $d^1$ double perovskite Ba$_2$NaOsO$_6$ 
\cite{Nikolov2026}.   
At lower temperature ($T_N \approx 18$ K) BMRO undergoes a second transition into a canted antiferromagnetic (cAFM) phase with wavevector $\vk = [0\ 0\ 1]$ and canting angle $\phi \sim 40^\circ$ relative to the $[1\ 1\ 0]$ crystallographic direction. 

Despite extensive experimental evidence for quadrupolar order, the microscopic origin of the observed phases remains an open question. An early theoretical study by Chen and coworkers proposed that the relevant physics could be captured by superexchange and cQQ interactions~\cite{Chen2010}. This framework was subsequently adopted by Svoboda and colleagues~\cite{Svoboda2021}; while both studies reproduced the AF\xs and cAFM orders, neither was able to stabilize the experimentally observed F\zs quadrupolar order. Subsequent work showed that retaining the full SOC GSM derived from the $t_{2g}$ manifold generates couplings between the AF\xs and F\zs moments, offering a possible explanation for the missing component~\cite{Kubo2023}. However, while experimentally the $z^2$ component is found to be sizable~\cite{Hirai2020_PRR}, the $z^2$  magnitude obtained in Ref.~\cite{Kubo2023} is minute as compared to that of the AF$x^2$-$y^2$ component. Moreover,  the approaches  of Refs.~\cite{Chen2010,Svoboda2021,Kubo2023} neglect vibronic interactions, which modify the ground-state wavefunctions through electron-phonon coupling~\cite{Iwahara2024}. Iwahara and colleagues~\cite{Iwahara2023} included both exchange and vibronic coupling within the \jeff manifold  and obtained the experimentally observed composite order assuming a  strong limit for the JT coupling and a selected set of model parameters.
More recently, Hart and colleagues identified a quartic coupling between neighboring \xs quadrupoles and an \zs-type distortion as a mechanism for stabilizing the F\zs component~\cite{Hart2026}, although its quantitative relevance remains to be established.
While 
first-principles density functional theory (DFT) calculations have elucidated the roles of electronic and structural degrees of freedom, particularly in the time-reversal-broken phase~\cite{Tehrani2021, Martinelli2026}, 
they
did not quanitatively capture the composite quadrupolar order in the paramagnetic phase.

In a previous work, we derived the IEI and vibronic couplings from first principles within the \jeff= 3/2 GSM,
%
capturing
the AF\xs order and the cAFM phase~\cite{Fioremosca2024b}. Here, we extend this approach by computing all relevant interactions, including contributions involving the excited \jeff=1/2 states and the  cQQ interaction. To our knowledge, although these contributions have been invoked in the past, their magnitudes have never been quantified from first principles. 

By fully including both  purely electronic and vibronic couplings we quantitatively reproduce the
%
%
composite
quadrupolar order
%
with a sizable $Fz^2$ component
and the accompanying structural distortions. 
Resolving the competing contributions of various coupling mechanisms and the role of \jeff=1/2 admixture we identify the microscopic origin of the antiferroic \xs and ferroic \zs orders and establish the relative energy scales of the interactions involved. Our results provide a unified microscopic description of the quadrupolar phase and a framework for understanding the emergence of multipolar order in double perovskites.

\vspace{0.2cm}
{\it Model Hamiltonian. ---} The effective low-energy Hamiltonian of BMRO acts in the full SOC $(J_{\mathrm{eff}}=3/2) \oplus (J_{\mathrm{eff}}=1/2)$ manifold. To represent local degrees of freedom within this manifold  we employ double-$J$ or "mixed" normal $O_{KQ}(JJ')$ and anomalous $\Omega_{KQ}(JJ')$ multipolar operators 
\cite{Blum_DM}, which separate the $J =3/2$ and $J'=1/2$ contributions explicitly (see End Matter for details). We introduce the compact notation $\mathcal{O}^{\nu}_{\mu}$ for these operators,
where 
the superscript $\nu \equiv \{JJ't\}$ labels the possible pairs of $J$ manifolds and the operator type $t$, defined as normal ($t=n$) and anomalous ($t=a$). The subscript  $\mu \equiv \{K, Q\}$ labels the multipolar rank and its projection. For $\nu = \{J,J,n\}$, the operators reduce to the conventional "single-$J$" multipoles $O_{KQ} (J)$~\cite{Santini2009}. 

Altogether, the effective Hamiltonian
contains 
five components
\beq
\label{Eq:all}
H = \sum_{\langle ij \rangle} \big( H_{ij}^{\rm IEI} + H_{ij}^{\rm cQQ} + H_{ij}^{\rm IEC} \big) + \sum_i \big(H_i^{\rm JT} + H_i^{\mathrm{SO}}\big), 
\eeq
with each 
term defined below.


The IEI term  between a pair of Re ions on bond $ij$ takes the general form
\beq
\label{Eq:iei}
H_{ij}^{\rm IEI}  = \sum_{\mu\mu'}\sum_{\nu \nu'} V_{\mu \mu'}^{\nu \nu'}(ij)  \mathcal{O}_{\mu}^{\nu} (i) \mathcal{O}_{\mu'}^{\nu'} (j),
\eeq
where  $V_{\mu \mu'}^{\nu \nu'}(ij)$ is the corresponding coupling constant.

The cQQ, IEC and JT terms involve only time-even quadrupoles. Let $\mathcal{N} = \{\frac{3}{2} \frac{3}{2} n, \frac{3}{2} \frac{1}{2} n \}$ denote the normal operator subspace and $\Gamma = \{ E_g, T_{2g} \}$ the subspace 
of quadrupoles
spanning the $E_g \oplus T_{2g}$ irreducible representations. Then the cQQ interaction is written as 
\beq
\label{Eq:qq}
H_{ij}^{\rm cQQ} = \sum_{\substack{\nu \nu' \in \mathcal{N}}} \sum_{\substack{\mu \mu' \in \Gamma}} I^{\nu \nu'}_{\mu \mu'} (ij) \mathcal{O}_{\mu}^{\nu}(i) \mathcal{O}_{\mu'}^{\nu'}(j),
\eeq
where $I^{\nu \nu'}_{\mu \mu'} (ij)$ denotes the electrostatic coupling. 

\begin{figure}[t]
  \centering
  \subfloat[\normalsize $q_{\mu = (2 0 )} =  3z^2-r^2$\label{fig:mode-eu}]{\octaEu}%
  \hspace{2mm}%
  \subfloat[\normalsize $q_{\mu = (2 2 )} = x^2-y^2$\label{fig:mode-ev}]{\octaEv}\\[-1.0em]
  \subfloat[]{\includegraphics[width=0.9\linewidth]{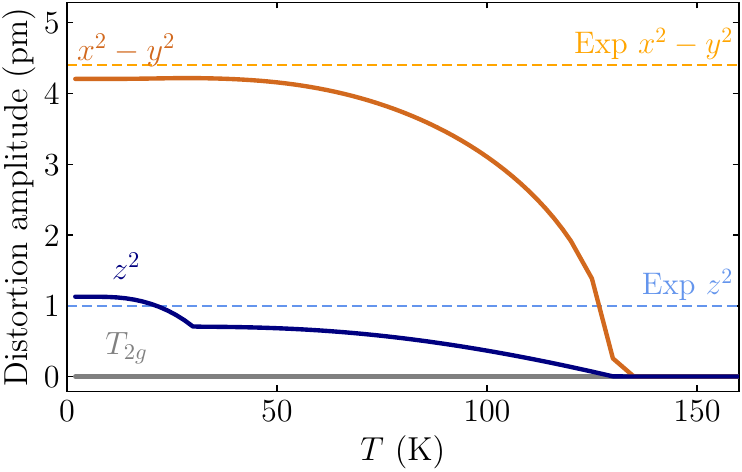}}
  \caption{
  $E_g$ Jahn--Teller (JT) distortions of the ligand octahedron:
    (a) the tetragonal \zs mode and (b) the orthorhombic \xs mode.
    (c) Calculated mean-field distortion amplitudes of the active JT modes as a function of temperature.  Below $T_q \sim $ 130 K both \zs and \xs 
    distortions appear, while below $T_N \sim $ 30 K the  \zs distortion further changes  in magnitude. Horizontal dashed lines are the experimental distortions measured at 6 K~\cite{Hirai2020_PRR}. }
  \label{fig:1}
\end{figure}

The vibronic part of the Hamiltonian contains both IEC and local JT contributions. The IEC between sites $i$ and $j$ is given by
\beq
\label{Eq:iec}
H_{ij}^{\rm IEC} = \sum_{\substack{\mu \mu' \in \Gamma}} \Phi_{\mu, \mu'} (ij) q_{\mu} (i) q_{\mu'} (j),
\eeq
where $\Phi_{\mu \mu'}(ij)$ denotes the irreducible force-constant matrix and $q_{\mu}$ are the corresponding irreducible distortion coordinates (See Fig.~\ref{fig:1}(a) and (b)). 

The local JT contribution can then be written as
\begin{align}
H_{i}^{\rm JT}  =  \sum_{\mu \in \Gamma} \frac{1}{2} \Bigg[ \frac{p^2_\mu (i)}{M}  + \Phi_{\mu \mu} q^2_{\mu} (i) \Bigg]  \label{Eq:jt} \\ - \sum_{\mu \in \Gamma} \sum_{\nu \in \mathcal{N}} g^{\nu}_{\Gamma} \mathcal{O}_\mu^\nu (i) q_{\mu} (i)  \nonumber,
\end{align}
where $p_\mu$ is the momentum conjugate to the distortion, $M$ is the reduced mass, $\Phi_{\mu\mu}$ is the diagonal on-site element of the irreducible force-constant matrix and $g_{\Gamma}^{\nu}$ the JT coupling constant coupling both single and double-$J$ quadrupoles (See end Matter for additional details). 

Finally, in the spin-orbit term $H^{\mathrm{SO}}_i$, the only non-zero matrix elements are the diagonal entries of the $J=1/2$ subblock equal to $3\lambda/2$.

\vspace{0.2cm}
{\it Methods. ---} We determine all couplings entering Eq.~\ref{Eq:all} from first principles using a combination of complementary approaches.
$H^{\mathrm{SO}}$ is  obtained 
within the full-charge self consistent DFT~\cite{Wien2k}+ dynamical mean field theory in the Hubbard-I (HI) approximation (DFT+HI)~\cite{Georges1996, Anisimov1997_1,Lichtenstein_LDApp,Aichhorn2016}. 
The IEI of Eq.~\ref{Eq:iei} are computed using the Force-Theorem in Hubbard-I (FT-HI) method of Ref.~\cite{Pourovskii2016} applied as a postprocessing of the DFT+HI paramagnetic electronic structure. The FT-HI  is implemented in the MagInt package~\cite{magint} and extended to account for mixed multipole operators. The details of the method and the resulting IEI can be found in the Supplemental Material (SM)~\cite{supplmat}. 
The cQQ couplings of Eq.~\ref{Eq:qq} are   obtained within a constrained DFT+HI framework
from intersite response to a local quadrupolar field 
as detailed in End Matter.  The JT coupling constants $g^\nu_\mu$ are computed within DFT+HI from the response of the corresponding electronic quadrupole moments to lattice distortions of the same symmetry. Finally, the IEC of Eq.~\ref{Eq:iec} and the on-site harmonic force constants entering Eq.~\ref{Eq:jt} have been already determined in Ref.~\cite{Fioremosca2024b} using a first-principles projective 
dynamical matrix approach~\cite{Giannozzi1991,Fioremosca2024b} described therein. 
Further details are provided in the SM S2~\cite{supplmat}.

To solve Eq.~\ref{Eq:all}, we employ a single site mean-field (MF) approximation, decoupling the IEI, cQQ, and IEC terms.  The resulting local vibronic problem is treated in a normalized product basis, $|\nu \rangle = |M\rangle \times |n \rangle$, where $|M \rangle$  belongs either to the $|\tfrac{3}{2},M\rangle$  manifold or to the  $|\tfrac{1}{2},M\rangle$ one and $|n \rangle$ denotes the phonon basis. Given that the  JT coupling to $T_{2g}$ distortions is almost three times smaller compared to the $E_g$ one, we restrict the phonon basis to the latter modes, such that $|n\rangle = |n_{z^2}, n_{x^2-y^2} \rangle$.
We have verified convergence with respect to the phonon basis and find that the quadrupole moments, MF transition temperatures, and structural distortions are fully converged for $n_{\rm tot} = n_{z^2} + n_{x^2-y^2}~\geq~4$. Additional details are provided in the SM S4~\cite{supplmat}.

\vspace{0.2cm}
{\it Quadrupolar order. ---} We first present results for $n_{\rm tot}=4$. BMRO undergoes two successive phase transitions:  a composite quadrupolar state, which is a superposition of the AF\xs order with propagation vector $\vk=[0\ 0\ 1]$ and ferroic F\zs one emerges at $T_q\sim130$ K,  followed by the onset of   cAFM order at $T_N\sim30$ K. In the cAFM phase, we obtain a canting angle $\phi\approx20^\circ$ and an ordered moment of $\approx0.36 \mu_B$ along the $[1\ 1\ 0]$ direction, in close agreement with the experimental value of $\approx0.3 \mu_B$. Remarkably, the calculation captures both the composite quadrupolar order preceding the magnetic transition and the accompanying structural distortion, with the latter in very good agreement with the values measured at 6 K by synchrotron X-ray diffraction (See Fig.~\ref{fig:1}(c))~\cite{Hirai2020_PRR}.

\begin{figure*}[ht]
  \begin{center}
  \includegraphics[width=\textwidth, clip=true]{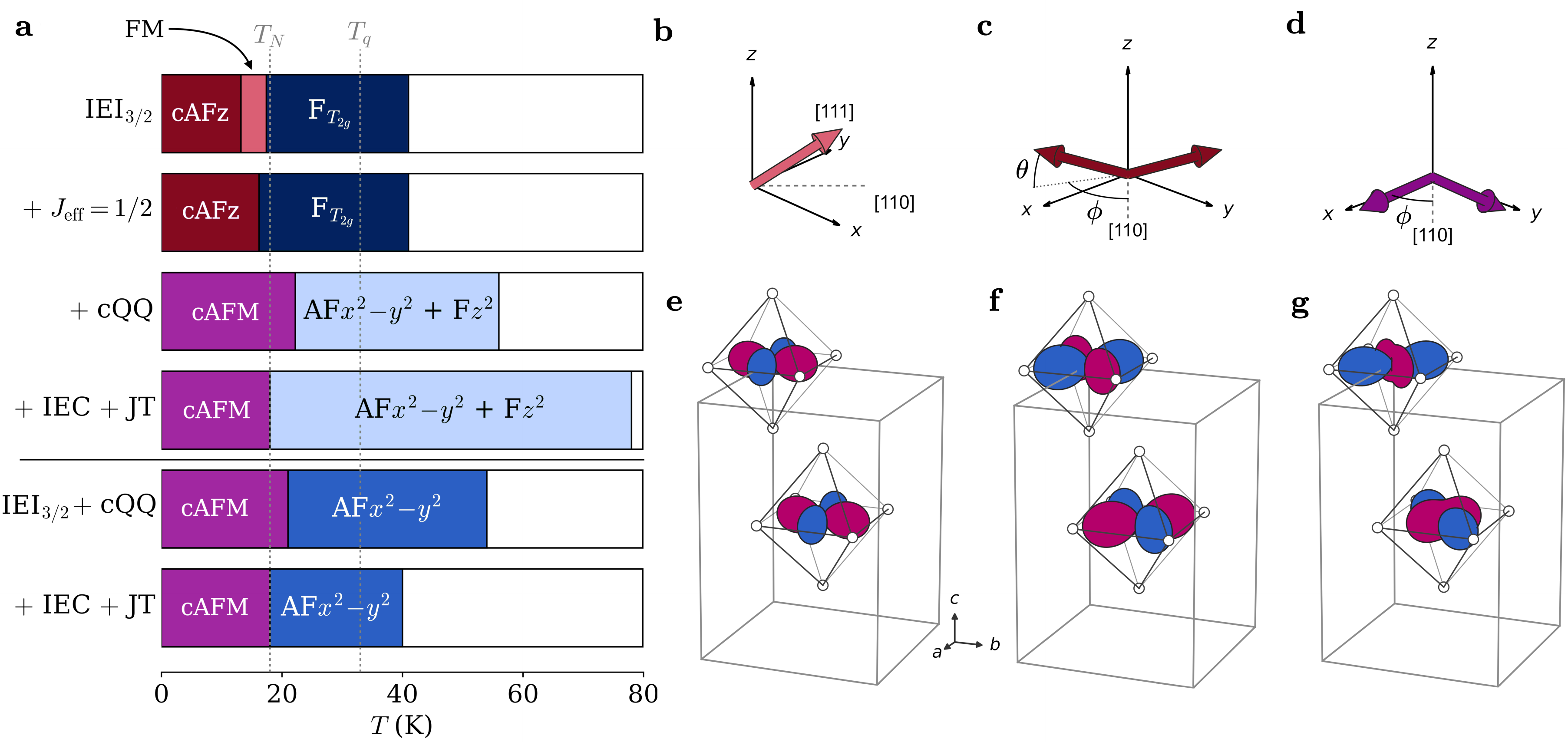}
    \end{center}
\caption{(a) Finite-temperature phase diagram of Eq.~\eqref{Eq:all} for different combinations of active interactions, obtained within single site mean field theory. (b)--(d) Magnetic dipolar orders: ferromagnetic with net moment along $[111]$ (FM), canted antiferromagnetic with the moments tilted out of the $(001)$ plane (cAFz), and canted antiferromagnetic (cAFM). (e)--(g) Quadrupolar orders: ferroquadrupolar order of $T_{2g}$ quadrupoles (F$_{T_{2g}}$), antiferroquadrupolar $x^2-y^2$ order (AF\xs), and coexisting AF\xs and ferroquadrupolar $z^2$ order (AF\xs+F\zs).}
\label{fig:2} 
\end{figure*}


As expected by the MF+HI approximation, the transition temperatures are substantially overestimated~\cite{Pourovskii2019, Pourovskii2021}. In particular, while the calculated $T_N$ is larger than the experimental value by about a factor of 1.6, in agreement with expectations for a MF overestimation, quadrupolar $T_q$ is overestimated by roughly a factor of four. 
The strong overestimation of $T_q$ 
could be explained by 
the formation of a glassy quadrupolar phase above  $T_q$ in real BMRO, as recently inferred for the isoelectronic double perovskite Ba$_2$NaOsO$_6$~\cite{Nikolov2026}. The glassy regime lies beyond the reach of single-site MF, but it is expected to feature lower total energy and entropy compared to the local-moment paramagnet resulting in the critical temperature for the order-glass transition being much lower than for the order-disorder one.

\vspace{0.2cm}
{\it Role of Individual Interactions. ---} We next examine the role of the individual interactions. Since the couplings entering Eq.~\ref{Eq:all} are determined independently, we can progressively add the different contributions and disentangle their effects. Figure~\ref{fig:2}~a summarizes this analysis. For clarity, the calculated transition temperatures in all cases are rescaled by the same multiplicative factor of 0.6, which is
chosen such as to match $T_N$ obtained with the full Hamiltonian (\ref{Eq:all}) to the experimental value.

We first consider the case of IEI within the isolated \jeff = 3/2 manifold (IEI$_{3/2}$), which exhibits three successive second-order phase transitions (see also Ref.~\cite{Fioremosca2024b}). The first is to a ferro-quadrupolar ordering of the $T_{2g}$ quadrupoles (F$_{T_{2g}}$; Figure~\ref{fig:2}e). It is followed by a transition to a ferromagnetic (FM) state with the net moment along $[1\ 1\ 1]$ (Figure~\ref{fig:2}b) and, at lower temperature, to a canted antiferromagnetic  (cAFz) phase (Figure~\ref{fig:2}c). We label this phase cAFz because the moments are tilted out of the $(001)$ plane by $\theta \approx 25^\circ$. Including the contributions of the \jeff=1/2 states (+\,\jeff=1/2) leaves the F$_{T_{2g}}$ transition essentially unchanged and only modestly alters the magnetic order. The intermediate FM phase is suppressed, and the low-temperature magnetic phase remains cAFz with a slightly smaller tilt, $\theta \approx 20^\circ$.

Adding the cQQ interactions (+\,cQQ) has a considerably stronger effect on both transitions. The cQQ coupling stabilizes the AF\xs order and induces a small F\zs component, with about 0.7\% of the AF\xs amplitude. Both the $T_q$ and $T_N$ temperatures shift upward, and the system now orders into a cAFM phase with canting angle $\phi \approx 28^\circ$.
While the weak F\zs component is consistent with the scenario proposed by Kubo and coworkers~\cite{Kubo2023}, it does not account for the sizable F\zs order observed experimentally. The decisive contribution instead comes from the vibronic couplings (+\,IEC\,+\,JT). In fact, with a converged phonon basis, the F\zs component becomes significant and is accompanied by lattice distortions in quantitative agreement with experiment. Crucially, our calculations deliver essentially the same result when the cQQ interactions are omitted. Thus, although cQQ interactions can generate a small F\zs component, they are neither necessary nor decisive for the emergence of the experimentally relevant F\zs order. The coexistence of AF\xs and F\zs quadrupole orders is illustrated in Fig.~\ref{fig:2}(g).

The AF\xs order alone, shown in Fig.~\ref{fig:2}f, can be stabilized by several combinations of interactions, including IEI$_{3/2}$\,+\,cQQ and IEI$_{3/2}$\,+\,IEC\,+\,JT~\cite{Fioremosca2024b} (See bottom panel of Fig.~\ref{fig:2}). By contrast, no F\zs component emerges within the reduced \jeff = 3/2 GSM, even when all other interactions are included. This indicates that the \jeff = 1/2 states are essential for the F\zs order.

\vspace{0.2cm}
{\it Origin of F\zs ---} The ferroic order of \zs, together with its associated JT distortion, originates from a linear-quadratic coupling between the single-$J$ \zs and the double-$J$ \xs quadrupoles. In Landau theory, this coupling corresponds to the free-energy term proportional to
$O_{z^2}\!\left(\tfrac{3}{2}\right) \left[ O_{x^2-y^2}\!\left(\tfrac{3}{2}\tfrac12\right) \right]^2$, as previously suggested in Ref.~\cite{Kubo2023}.
The reason is that it is only the mixed $O_{x^2-y^2}(\frac{3}{2}\frac{1}{2})$ quadrupole, squared and projected to the \jeff= 3/2 manifold, that yields a term proportional to the \jeff= 3/2 $z^2$ quadrupole. Squaring the single-$J$ $O_{x^2-y^2}(\frac{3}{2})$ quadrupole gives a multiple of the identity and does not lead to any coupling. 
Correspondingly, a sizable F\zs order at the MF level requires an appreciable mixed $x^2-y^2$ term, which, as shown above, arises only once the vibronic interactions are included.


The reason is that the on-site JT coupling  (Eq.~\ref{Eq:jt}) links the mixed  $O_{x^2-y^2}(\frac{3}{2}\frac{1}{2})$ quadrupole to the on-site distortion $q_{x^2-y^2}$, therefore inducing the required sizable $O_{x^2-y^2}(\frac{3}{2}\frac{1}{2})$ term once the AF\xs order and the corresponding distortion appear. This term induces the \zs component as a secondary order parameter,
which in turn drives the corresponding structural distortion,
the latter then induces  the order of the mixed \zs quadrupole.
Without the JT coupling, the mixed $x^2-y^2$ term arises solely from the comparatively weak intersite interactions, so the purely electronic IEI\,+\,cQQ model yields only a negligible F\zs component even when the \jeff$=1/2$ states are included.

\begin{figure}[t]
  \begin{center}
  \includegraphics[width=\columnwidth, clip=true]{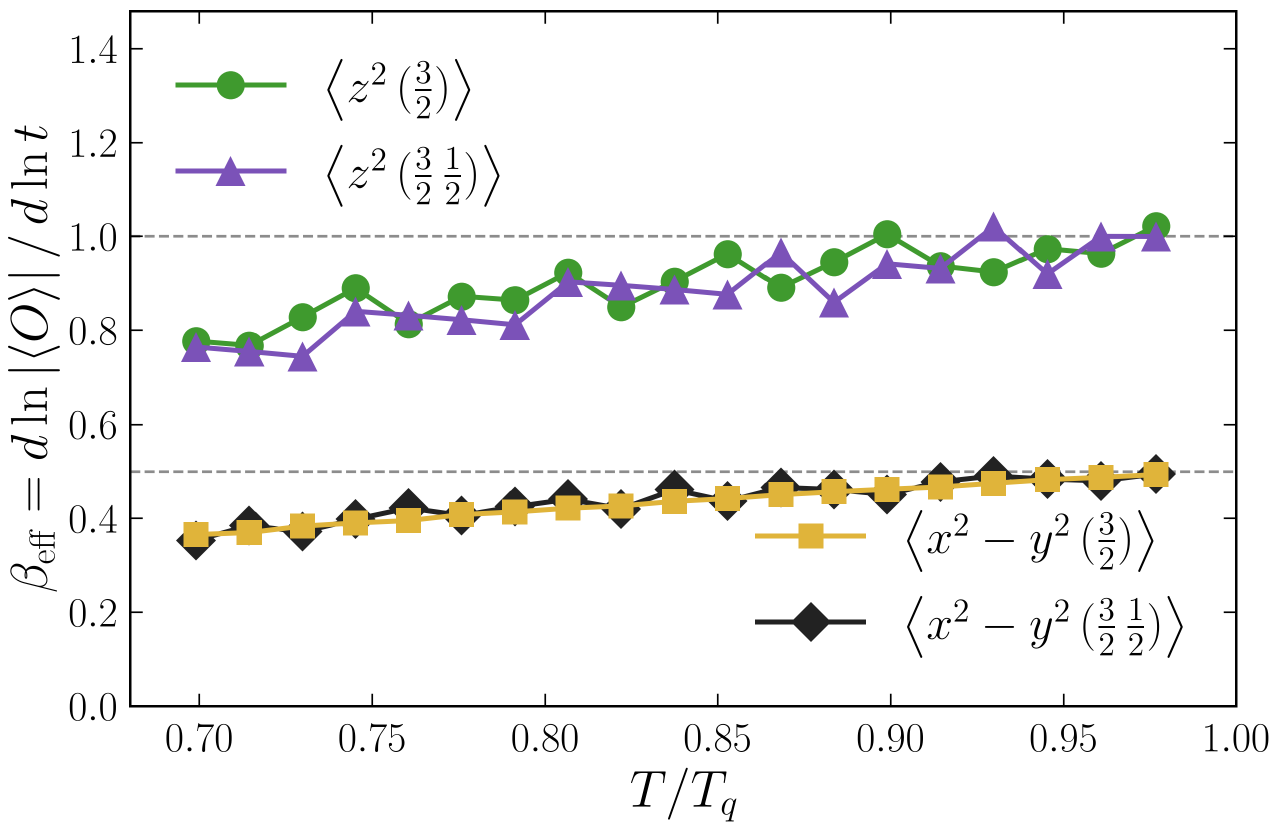}
    \end{center}
\caption{Effective critical exponent $\beta_{\mathrm{eff}}=d\ln|\langle O\rangle|/d\ln t$ with $t = 1 - T/T_q$ for the primary AF\xs and secondary F\zs order parameters in mean field. Dashed lines mark the values $\beta = 1/2$ and $1$. 
}
\label{fig:3} 
\end{figure}

This mechanism is confirmed 
%
by the evolution of the critical exponents of the  $x^2-y^2$ and  $z^2$ quadrupole order parameters vs temperature
that we calculated 
in mean field at $T$ approaching $T_q$ (See Figure~\ref{fig:3}).  Both $J=3/2$ and mixed quadrupoles  exhibit the same critical exponent for a given projection, with the exponent of $z^2$ being twice larger than that of $x^2-y^2$.
 Moreover these quadrupoles behave differently as a function of the phonon cutoff (See SM~\cite{supplmat} S4). The primary order parameter $O_{x^2-y^2}(\frac{3}{2})$  is already converged at $n_{\mathrm{ph}} = 2$, whereas the $O_{z^2}(\frac{3}{2})$ and associated distortion continues to evolve with $n_{\mathrm{ph}}$, tracking the mixed primary order parameter $O_{x^2-y^2}(\frac{3}{2}\frac{1}{2})$.

\vspace{0.2cm}
{\it Conclusion. ---} To identify the origin of the composite quadrupolar order in \bmro, we computed the superexchange, electrostatic quadrupole-quadrupole, and vibronic interactions from first principles, including the effect of the excited \jeff=1/2 states through a double-$J$ multipolar-operator formalism, and solved the resulting effective Hamiltonian. The calculations reproduce the composite order, ferroic in \zs and antiferroic in \xs, and the associated structural distortions agree quantitatively with experiment. Both the \jeff=1/2 states and the vibronic coupling are essential for a sizable ferroic \zs order, which we show to be a secondary order parameter induced by the antiferroic \xs order. In contrast to earlier proposals, classical quadrupole-quadrupole interactions play a negligible role. 
While our theory reproduces all order parameters in good agreement with experiment and overestimates $T_N$ only by the usual MF amount, the predicted quadrupolar transition temperature is far too high to be accounted for by the same MF effect. This discrepancy hints at a different nature of the quadrupolar transition in BMRO, possibly involving a glassy Jahn-Teller phase inferred from recent experiments on related systems~\cite{Nikolov2026}.

These results revise the current picture of quadrupolar order in 5$d^1$ double perovskites and raise the questions on isoelectronic compounds, such as the antiferromagnet Ba$_2$LiOsO$_6$, which can develop different multipolar orders. Calculations at different volumes using the full effective Hamiltonian do not significantly alter the ordering landscape (see End Matter), indicating that changes in lattice volume alone cannot account for this difference. 

More generally, the formalism developed here applies to other spin-orbit-coupled $d^n$ systems in which exchange, vibronic coupling, and excited multiplets compete on comparable energy scales, and can help identify the mechanisms that drive their multipolar orders.

\hspace{0.1cm}

\begin{acknowledgements}
This research was supported by the Austrian Science Fund (FWF) project DOI: 10.55776/PAT2121825.
D.~F.~M. thanks the computational facilities of the Austrian Scientific Computing (ASC). L.~V.~P. acknowledges the support by CNRS through the Tremplin@Physique 2025 program and  by the  CPHT computer team. 
\end{acknowledgements}

\nocite{hubbard_1,Kresse1993,Kresse1996,
phonopy-phono3py-JPSJ,Bersuker1989,brooks1997density,
Fioremosca2021,Fioremosca2025,Teruki2026,Englman70,
Gehring_1975,Polinger2009,Rotter2004}

\bibliography{bibliography}

\clearpage

\section{End Matter}
\begin{figure*}[ht]
  	\begin{centering}
  	\includegraphics[width=\textwidth]{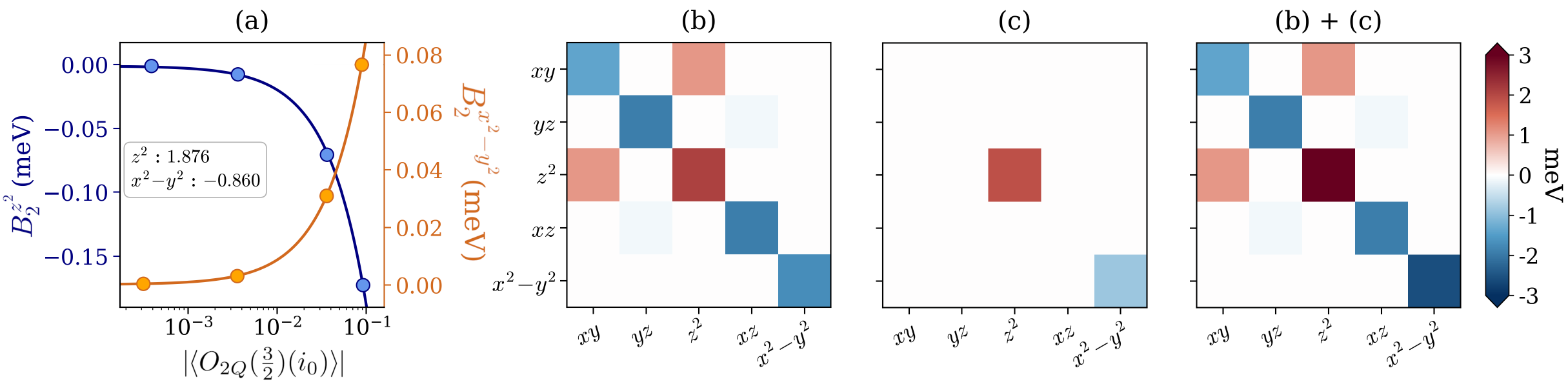} 
  		\par\end{centering}
  	\caption{(a) Mean quadrupolar fields $B_2^{z^2}$ (blue, left) and $B_2^{x^2-y^2}$ (orange, right) as functions of the quadrupole expectation value $\langle O_{2Q} (\frac{3}{2}) (i_0) \rangle$, shown on a logarithmic scale. Circles: DFT+HI data; lines: linear fits. (b) Intersite exchange interactions coupling QQ terms and (c) cQQ interactions for $\vR = [0.5, 0.5, 0]$, and (b+c) their sum.}
\label{fig:end_matt_1}
\end{figure*}


\vspace{0.2cm}
{\it Generalized multipole operators. ---} Several definitions of operators spanning the $\left[\tilde{l}=1\right] \times \left[S=1/2\right]$ space are available in the literature. In particular, Refs.~\cite{Chen2010,Svoboda2021,Kubo2023} express them as direct products of $\tilde{l}=1$ orbital and $S=1/2$ one-electron spin operators, while Refs.~\cite{Wang2017,Fan2026} construct a set of irreducible spherical tensors from these products. Neither approach explicitly separates the $J=3/2$ and $J=1/2$ contributions in the resulting operator set. To single out the effect of the $J=1/2$ admixture, we instead construct the operators from double-$J$ spherical tensors~\cite{Blum_DM}, as detailed in the Supplemental Material (SM)~\cite{supplmat}. We note that this construction is quite general and applies equally well to many-electron $J$ multiplets.

The resulting double-$J$ operators employed in Eqs.~(\ref{Eq:all})--(\ref{Eq:jt}) 
are $\mathcal{O}^{\nu}_{\mu}\equiv\mathcal{O}^{tJJ'}_{KQ}$, where $\mu=(K,Q)$ denotes the rank $K$ and component $Q$, and $\nu=(t,J,J')$ the operator type $t$ and the pair of $J$ manifolds it connects:

\beq
  \mathcal{O}^{tJJ'}_{KQ}=
  \begin{cases}
    O_{KQ}(JJ'),      & t=n,\\
    \Omega_{KQ}(JJ'), & t=a.
  \end{cases}
\eeq

For $J=J'$ these reduce to the conventional single-$J$ multipoles $O_{KQ}(J)$. For $J\neq J'$, each $(K,Q)$ with $|J-J'|\le K\le J+J'$ admits two independent Hermitian operators acting between the two manifolds: the \emph{normal} multipoles $O_{KQ}(JJ')$, which have the same time-reversal parity, $(-1)^K$, as single-$J$ multipoles of equal rank, and the \emph{anomalous} (toroidal) multipoles $\Omega_{KQ}(JJ')$, which have the opposite parity, $(-1)^{K+1}$. 

The operators $O_{KQ}(J)$, $O_{KQ}(J')$, $O_{KQ}(JJ')$, and $\Omega_{KQ}(JJ')$ together form a complete set in the $J\oplus J'$ space and orthonormal under the trace inner product, so that any operator acting in this space can be expanded with them. For $J=3/2$ and $J'=1/2$ they are $6\times6$ matrices. The single-$J$ sector comprises sixteen $O_{KQ}(\tfrac{3}{2})$ (monopole, dipoles, quadrupoles, and octupoles, $K=0$--$3$) and four $O_{KQ}(\tfrac{1}{2})$ (monopole and dipoles, $K=0,1$). The mixed sector ($K=1,2$) comprises eight normal operators $O_{KQ}(\tfrac{3}{2}\tfrac{1}{2})$, i.e., magnetic dipoles and electric quadrupoles, and eight anomalous operators $\Omega_{KQ}(\tfrac{3}{2}\tfrac{1}{2})$, i.e., time-reversal-even dipoles and time-reversal-odd quadrupoles. 

For clarity,  Eqs.(\ref{Eq:jt}) defines two couplings $g_\Gamma (\tfrac{3}{2}\tfrac{1}{2})$ and $ g_\Gamma(\tfrac{3}{2})$ and it can be explicitly written, omitting the $(i)$ on-site label, as 

\begin{align}
H^{\rm JT}  =  & \sum_{Q \in \Gamma}  \Bigg[ \frac{p^2_Q}{2M}  + \frac{\Phi_{Q Q}}{2} q^2_{Q}   \label{Eq:jt} \\ & - \Big[ g_{\Gamma} (\tfrac{3}{2}\tfrac{1}{2}) O_{KQ} (\tfrac{3}{2}\tfrac{1}{2})   +  g_{\Gamma} (\tfrac{3}{2}) O_{KQ} (\tfrac{3}{2})\Big] q_{Q} \Bigg] \nonumber .
\end{align}

\vspace{0.2cm}
{\it Classical quadrupole--quadrupole interactions.---} We determine the classical quadrupole--quadrupole (cQQ) couplings from fully charge self-consistent DFT+HI calculations using a trial-field supercell approach. 
In this approach we carry out DFT+HI calculations for a BMRO supercell introducing a quadrupole field on a \emph{trial} Re site $i_0$ and  measuring a response to this field on another, \emph{measurement}, Re site $i$. The field on the trial site induces a corresponding local quadrupole, which in turn generates
a quadrupolar field on the measurement site. 
The approach takes advantage of the principal limitation of the Hubbard-I approximation, namely the neglect of the hybridization function and, consequently, of kinetic intersite exchange. As a result, the effective quadrupolar field acting on the measurement site contains only the classical electrostatic contribution, since the exchange contribution is suppressed.

To the
trial site we apply a static single-particle field of definite $E_g$ or
$T_{2g}$ symmetry,
\beq
\label{eq:HQ}
\hat H_Q = \Delta_Q \sum_{\alpha\beta} \Big[O_{2Q}(\tfrac{3}{2})\Big]_{\alpha\beta}\,
c^\dagger_{i_0\alpha} c^{\phantom{\dagger}}_{i_0\beta},
\eeq
where $\alpha,\beta$ label 
the six 
states of the (\jeff=3/2)$\oplus( $\jeff=1/2)
manifold. 
No external field acts at the measurement site, so any quadrupolar field acquired is generated by the polarized trial site and its images generated due to periodic boundary conditions (computational details in the SM~\cite{supplmat} S2).

From each self-consistent 
supercell calculation 
we extract the induced quadrupoles at the trial site $\langle O_{2Q} (\frac{3}{2}) (i_0)  \rangle = \mathrm{Tr} [O_{2Q} (\frac{3}{2}) \rho ]$, where $\rho $ is the local density matrix in the (\jeff=3/2)$\oplus( $\jeff=1/2) space,
and the mean quadrupolar fields at the measurement site, $B_2^Q (i) = \mathrm{Tr} [O_{2Q} (\frac{3}{2}) \epsilon (i)]$, where $\epsilon (i)$ are the on-site $t_{2g}$ level positions in the same space.

By restricting the cQQ intersite coupling of Eq.~\ref{Eq:qq} to the \jeff$=3/2$ subspace defined by $\nu = \nu' = \left(\tfrac32 \tfrac32 n\right) \equiv 3/2$ and $K = K' = 2$ and applying the mean field decoupling we find 
\beq
B_2^Q (i) = \sum_{j\neq i}\sum_{Q'} I^{3/2}_{QQ'}\langle O_{2Q'} (\tfrac{3}{2})(j) \rangle \ .
\eeq
so that the cQQ coupling constant is the linear response coefficient 
\beq
  I^{3/2}_{QQ'} = \frac{1}{N_{\rm im}}\left.\frac{\partial B_2^Q (i)} {\partial\langle O_{2Q} (\frac{3}{2}) (i_0) \rangle}\right|_{\Delta_Q\to0},
\label{eq:V}
\eeq
where $N_{\mathrm{im}} = 4$ is the number of   equivalent periodic images of $i_0$, with $I>0$ corresponding to an antiferroquadrupolar coupling. $I^{3/2}_{QQ'}$ is obtained from a linear regression (See Figure~\ref{fig:end_matt_1} a). 
We restrict the cQQ to the $E_g$ channel due to the negligible effect of the quadrupolar field on the $T_{2g}$ in DFT+HI. As a result, we obtain $I^{3/2}_{00} \equiv I_{3/2}^{z^2}=1.88$~meV and $I^{3/2}_{22} \equiv I_{3/2}^{x^2-y^2}=-0.86$~meV for the   bond $\vR = [0.5,\ 0.5,\ 0]$, which are then added to the IEI at the quadrupolar channel, as shown graphically in Figure~\ref{fig:end_matt_1} (b-c), where we display the QQ interactions coming from the IEI, the computed cQQ interactions and their sum for this representative bond. 

To extend the cQQ coupling to the full SOC manifold,
we use the fact that the cQQ interaction within \jeff=3/2 is just a projection of the  cQQ coupling in the full $t_{2g}$ shell. 
The corresponding $t_{\mathrm{2g}}$ quadrupole operators transformed into the (\jeff= 3/2)$\oplus( $\jeff= 1/2) basis read
\beq
  O^{t_{2g}}_{2Q}=\sum_{\nu}c_{\nu}\,\mathcal{O}^{\nu}_{2Q} = O_{2Q}(\tfrac32)-\sqrt{2}\,O_{2} (\tfrac32\tfrac12),
  \label{eq:Ot2g}
\eeq
i.e.\ $c_{3/2}=1$ and $c_{3/2\,1/2}=-\sqrt{2}$, with no \jeff$=1/2$ component (see SM~\cite{supplmat} S2). By replacing the \jeff$=3/2$ projected operators $O_{2Q}(\tfrac32)$ with $O^{t_{2g}}_{2Q}$ one obtains a relation between the cQQ interactions in the corresponding spaces 
\beq
   I^{\nu\nu'}_{2Q2Q'} = c_\nu c_{\nu'} I^{3/2}_{QQ'} \ ,
\eeq
or equivalently  
\beq
  \begin{aligned}
  I^{\frac32,\frac32\frac12}_{2Q,2Q'} = I^{\frac32\frac12,\frac32}_{2Q,2Q'} &= -\sqrt{2}\,I^{3/2}_{QQ'},\\
  I^{\frac32\frac12,\frac32\frac12}_{2Q,2Q'} &= 2\,I^{3/2}_{QQ'},
  \end{aligned}
  \label{eq:Iup_explicit}
\eeq
with the shorthand $\tfrac32\equiv\left(\tfrac32\tfrac32n\right)$ and $\tfrac32\tfrac12\equiv\left(\tfrac32\tfrac12n\right)$.

\begin{figure}[ht]
  	\begin{centering}
  	\includegraphics[width=0.95\columnwidth]{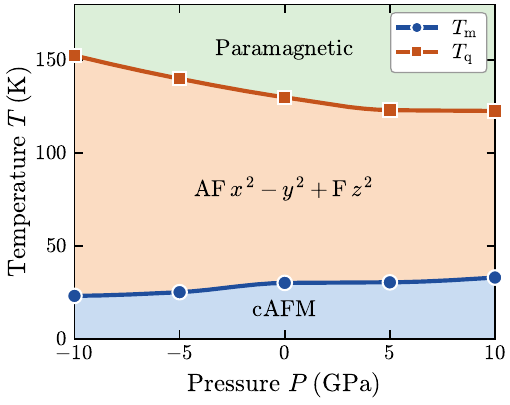} 
  		\par\end{centering}
  	\caption{Pressure--temperature phase diagram of \bmro\ computed with the IEI, the JT couplings, the IEC, and the cQQ interactions in the full SOC space. Squares mark the quadrupolar transition temperature $T_{\mathrm{q}}$ and circles the N\'eel temperature $T_{\mathrm{N}}$. Temperatures are not rescaled.} 
  	\label{fig:end_matt_2} 
\end{figure}

\vspace{0.2cm}
{\it Effect of pressure} To connect the physics of \bmro\ to isoelectronic double perovskites and to pressure experiments~\cite{Yammamura2006,Hirai2021_JPCM,Barbosa2022,Ishikawa2021,Arima2022}, we computed the IEI, the JT couplings, the IEC, and the cQQ interactions in the full SOC  space as a function of pressure $P$ between $-10$ and  $10$~GPa in steps of 5 GPa. 
The resulting phase diagram (Fig.~\ref{fig:end_matt_2}) shows that both the AF$x^2-y^2$+F$z^2$ quadrupolar order and the canted antiferromagnetic (cAFM) order persist over the entire pressure range.
The quadrupolar transition temperature $T_{\mathrm{q}}$ show a monotonic decrease with increasing $P$, while the N\'eel temperature $T_{\mathrm{N}}$ remains almost constant. This picture is broadly similar to our earlier results obtained without the cQQ interaction and without mixed multipolar IEI and JT couplings (See Fig.~2 of Ref.~\cite{Fioremosca2024b}). However, no collinear AFM phase predicted before~\cite{Fioremosca2024b} to appear close to $T_N$ upon compression  emerges  
 and $T_{\mathrm{N}}$ exhibits a markedly weaker dependence on pressure.

\clearpage

\end{document}


\title{Supplementary material for 
'Origin of quadrupolar order in the 5d$^1$ double perovskite Ba$_2$MgReO$_6$'}

\author{Dario Fiore Mosca}
\address{University of Vienna, Faculty of Physics, Center for Computational Materials Science, Kolingasse 14-16, 1090, Vienna, Austria}

\author{Leonid V. Pourovskii}
\address{CPHT, CNRS, \'Ecole polytechnique, Institut Polytechnique de Paris, 91120 Palaiseau, France}
\address{Coll\`ege de France, Université PSL, 11 place Marcelin Berthelot, 75005 Paris, France}

\date{\today}	
\maketitle

\makeatletter
\renewcommand{\l@subsubsection}[2]{}
\makeatother

\renewcommand{\thesection}{S\arabic{section}}

\tableofcontents

\section{Generalized multipolar operators}\label{sec:ops}

In order to include multipoles formed by both the $J=3/2$ ground state quadruplet and the $J=1/2$ excited doublet as well as their mixture we introduce double$-J$ multipolar operators~\cite{Blum_DM} as follows
\beq\label{eq:Tmixed}
T_{KQ}(JJ')=\sum_{MM'}(-1)^{J-M}(2K+1)^{1/2}
\begin{pmatrix}
J' & J & K \\
M' & -M & Q
\end{pmatrix}|JM\rangle\langle J'M'|.
\eeq

In the case of $J=J'$ they reduce to the standard definition for a single $J$ shell, see, e.~g., Ref.~\onlinecite{Santini2009}. For $J \ne J'$ the spherical tensor operators $T_{KQ}(JJ')$ have the integer rank $|J-J'| \le K \le |J+J'|$, with the standard  rule for the projection $Q=-K,-K+1,\dots, K$.
Moreover, the action of the time reversal operator $\Theta$ on $T_{KQ}(JJ')$ is given by 
$$
\Theta T_{KQ}(JJ') \Theta^{-1}=(-1)^{K-Q}T_{K\overline{Q}}(JJ'),
$$
where $\overline{Q}\equiv -Q$.

Introducing the adjoint $T^{\dagger}_{KQ}(JJ')$ of an operator $T_{KQ}(JJ')$ one can also show~\cite{Blum_DM} that
$$
T^{\dagger}_{KQ}(JJ')=(-1)^{J-J'+Q}T_{K\overline{Q}}(J'J),
$$
and that the operators $T_{KQ}(JJ')$  form an orthonormal set with their adjoints:
\beq\label{eq:orth}
\mathrm{Tr}\left[T_{KQ}(JJ')\cdot T^{\dagger}_{K'Q'}(JJ')\right]=\delta_{KK'}\delta_{QQ'}.
\eeq

Using the adjoints we introduce the following Hermitian operators for $J\ne J'$:
\beq\label{eq:opO}
\tilde{O}_{KQ}(JJ')=\frac{1}{\sqrt{2}}\left[T_{KQ}(JJ')+T_{KQ}^{\dagger}(JJ')\right]
\eeq
\beq\label{eq:opOmega}
\tilde{\Omega}_{KQ}(JJ')=\frac{i}{\sqrt{2}}\left[T_{KQ}(JJ')-T_{KQ}^{\dagger}(JJ')\right]
\eeq 

One may notice that the action of time reversal on the operators (\ref{eq:opO}) and (\ref{eq:opOmega}) flips their projection $Q$ and adds a $Q$-dependent phase:
$$
\Theta \tilde{O}_{KQ}(JJ')\Theta^{-1}=(-1)^{K-Q}\tilde{O}_{K\overline{Q}}(JJ'),
$$
$$
\Theta \tilde{\Omega}_{KQ}(JJ')\Theta^{-1}=(-1)^{K-Q+1}\tilde{\Omega}_{K\overline{Q}}(JJ').
$$
In order to obtain double-$J$ operators with definite parity under time reversal we thus combine the operators (\ref{eq:opO}) and (\ref{eq:opOmega}) to define new operators $O_{KQ}(JJ')$ as follows:
\beq\label{eq:opON_pQ}
O_{KQ}(JJ')=\frac{1}{\sqrt{2}}\left[(-1)^Q\tilde{O}_{KQ}(JJ')+\tilde{O}_{K\overline{Q}}(JJ')\right]
\eeq
\beq\label{eq:opON_nQ}
O_{K\overline{Q}}(JJ')=\frac{1}{\sqrt{2}}\left[(-1)^Q\tilde{\Omega}_{KQ}(JJ')-\tilde{\Omega}_{K\overline{Q}}(JJ')\right],
\eeq
and also use analogous definitions for the operators $\Omega_{KQ}(JJ')$:
\beq\label{eq:opOmegaN_pQ}
\Omega_{KQ}(JJ')=\frac{1}{\sqrt{2}}\left[(-1)^Q\tilde{\Omega}_{KQ}(JJ')+\tilde{\Omega}_{K\overline{Q}}(JJ')\right]
\eeq
\beq\label{eq:opOmegaN_nQ}
\Omega_{K\overline{Q}}(JJ')=\frac{1}{\sqrt{2}}\left[(-1)^Q\tilde{O}_{KQ}(JJ')-\tilde{O}_{K\overline{Q}}(JJ')\right],
\eeq
where $Q>0$ in the Eqs.~\ref{eq:opON_pQ}-\ref{eq:opOmegaN_nQ} above. The operators for $Q=0$ are unchanged:
\beq\label{eq:opOmegaN_0Q}
O_{K0}(JJ')=\tilde{O}_{K0}(JJ')
\eeq
\beq\label{eq:opOmegaN_0Q}
\Omega_{K0}(JJ')=\tilde{\Omega}_{K0}(JJ')
\eeq

One can verify that the action of time reversal on those operators (\ref{eq:opON_pQ}-\ref{eq:opOmegaN_0Q}) is given by
\beq\label{eq:TrevO}
\Theta O_{KQ}(JJ')\Theta^{-1}=(-1)^{K}O_{KQ}(JJ'),
\eeq
\beq\label{eq:TrevP}
\Theta \Omega_{KQ}(JJ')\Theta^{-1}=(-1)^{K+1}\Omega_{KQ}(JJ'),
\eeq
and that they form an orthonormal set of operators
{\setlength{\jot}{7px} 
\begin{flalign}\label{eq:Tr}
\mathrm{Tr}\left[O_{KQ}(JJ')\cdot O_{K'Q'}(JJ')\right]=\delta_{KK'}\delta_{QQ'} \nonumber\\\mathrm{Tr}\left[\Omega_{KQ}(JJ')\cdot \Omega_{K'Q'}(JJ')\right]=\delta_{KK'}\delta_{QQ'} \\\mathrm{Tr}\left[O_{KQ}(JJ')\cdot \Omega_{K'Q'}(JJ')\right]= 0 \nonumber
\end{flalign}}

Hermitian operators for a single $J-$shell,  $O_{KQ}(JJ) \equiv O_{KQ}(J)$, are formed in the standard way as described before~\cite{Santini2009}:
\begin{align}\label{eq:O_singJ}
O_{KQ}(J)=\frac{1}{\sqrt{2}}\left[(-1)^QT_{KQ}(JJ)+T_{K\overline{Q}}(JJ)\right] \\
O_{K\overline{Q}}(J)=\frac{i}{\sqrt{2}}\left[T_{KQ}(JJ)-(-1)^QT_{K\overline{Q}}(JJ)\right] \\
O_{K0}(J)=T_{K0}(JJ) 
\end{align}
where $Q > 0$. Their transformation under time reversal is the same as that of the $O_{KQ}(JJ')$ operators and is given by Eq.~\ref{eq:TrevO}.

All the operators -- the double-$J$ ones introduced in Eqs.~\ref{eq:opON_pQ}-\ref{eq:opOmegaN_0Q}, as well as the single-$J$ operators -- are Hermitian matrices of the dimension $2(J+J')+2$. Only the diagonal block corresponding to a given multiplet $J$ is non-zero in the single-$J$ operators. For the mixed operators with $J\ne J'$  only two off-diagonal blocks are non-zero. Therefore, the single-$J$ and double-$J$ operators are always orthogonal: $$ 
\mathrm{Tr}\left[O_{KQ}(JJ')\cdot O_{K'Q'}(J)\right]=0; \quad
\mathrm{Tr}\left[\Omega_{KQ}(JJ')\cdot O_{K'Q'}(J)\right]=0.
$$

Altogether, the single-$J$ operators $O_{KQ}(J)$ and $O_{KQ}(J')$ together with the double-$J$ operators  $O_{KQ}(JJ')$ and $\Omega_{KQ}(JJ')$ form a complete orthonormal set in the $J \oplus J'$ space, in which any operator acting in this space can be expanded.

In the case of $J=3/2$ and $J'=1/2$, the multipolar operators are thus $6 \times 6$ matrices. The single-$J$ operator set comprises the $J=3/2$ monopole, dipoles, quadrupoles, and octupoles, as well as the $J=1/2$ monopole and dipoles. The double-$J$ (mixed) operator set comprises the magnetic dipoles $O_{1Q}(\frac{3}{2}\frac{1}{2})$ and electrical quadrupoles $O_{2Q}(\frac{3}{2}\frac{1}{2})$ as well as the "anomalous" electric dipoles $\Omega_{1Q}(\frac{3}{2}\frac{1}{2})$ and magnetic quadrupoles $\Omega_{2Q}(\frac{3}{2}\frac{1}{2})$. The reason for this classification of mixed operators is due to their properties under time reversal summarized Table~\ref{tab:TR}.

Therefore, the full set of multipoles for the $J \oplus J'$ space consists of sixteen single-$J$ $O_{KQ}(\frac{3}{2})$, four single-$J$ $O_{KQ}(\frac{1}{2})$, eight $O_{KQ}(\frac{3}{2}\frac{1}{2})$, and eight $\Omega_{KQ}(\frac{3}{2}\frac{1}{2})$ operators. Altogether, there are thus 36 multipoles that fully determine the $6\times 6$  Hermitian density matrix of the $(J=3/2)\oplus(J'=1/2)$ manifold.

As an example, we give below the explicit form of the $z^2$ and $x^2-y^2$ quadrupole operators, which play the key role in BMRO quadrupolar order. The corresponding single-$J$ operators read
$$
O_{z^2}\left(\tfrac{3}{2}\right)\equiv O_{20}\left(\tfrac{3}{2}\right)=\begin{pmatrix}
   1/2 & 0 & 0 & 0 & 0 & 0 \\
   0 & -1/2 & 0 & 0 & 0 & 0 \\
   0 & 0 & -1/2 & 0 & 0 & 0 \\
   0 & 0 & 0 & 1/2 & 0 & 0 \\
   0 & 0 & 0 & 0 & 0 & 0 \\
   0 & 0 & 0 & 0 & 0 & 0 \\   
\end{pmatrix}, \quad 
O_{x^2-y^2}\left(\tfrac{3}{2}\right)\equiv O_{22}\left(\tfrac{3}{2}\right)=\begin{pmatrix}
   0 & 0 & 1/2 & 0 & 0 & 0 \\
   0 & 0 & 0 & 1/2 & 0 & 0 \\
   1/2 & 0 & 0 & 0 & 0 & 0 \\
   0 & 1/2 & 0 & 0 & 0 & 0 \\
   0 & 0 & 0 & 0 & 0 & 0 \\
   0 & 0 & 0 & 0 & 0 & 0 \\   
\end{pmatrix},
$$
and the double-$J$ electrical quadrupoles for the same projections read
$$
O_{z^2}\left(\tfrac{3}{2}\tfrac{1}{2}\right)\equiv O_{20}\left(\tfrac{3}{2}\tfrac{1}{2}\right)=\begin{pmatrix}
   0 & 0 & 0 & 0 & 0 & 0 \\
   0 & 0 & 0 & 0 & 1/2 & 0 \\
   0 & 0 & 0 & 0 & 0 & -1/2 \\
   0 & 0 & 0 & 0 & 0 & 0 \\
   0 & 1/2 & 0 & 0 & 0 & 0 \\
   0 & 0 & -1/2 & 0 & 0 & 0 \\   
\end{pmatrix}, \quad 
O_{x^2-y^2}\left(\tfrac{3}{2}\tfrac{1}{2}\right)\equiv O_{22}\left(\tfrac{3}{2}\tfrac{1}{2}\right)=\begin{pmatrix}
   0 & 0 & 0 & 0 & 0 & -1/2 \\
   0 & 0 & 0 & 0 & 0 & 0 \\
   0 & 0 & 0 & 0 & 0 & 0 \\
   0 & 0 & 0 & 0 & 1/2 & 0 \\
   0 & 0 & 0 & 1/2 & 0 & 0 \\
   -1/2 & 0 & 0 & 0 & 0 & 0 \\   
\end{pmatrix},
$$
where the $|JM\rangle$ basis is ordered as
\[
\left\{
\left|\tfrac{3}{2},-\tfrac{3}{2}\right\rangle,
\left|\tfrac{3}{2},-\tfrac{1}{2}\right\rangle,
\left|\tfrac{3}{2},\tfrac{1}{2}\right\rangle,
\left|\tfrac{3}{2},\tfrac{3}{2}\right\rangle,
\left|\tfrac{1}{2},-\tfrac{1}{2}\right\rangle,
\left|\tfrac{1}{2},\tfrac{1}{2}\right\rangle
\right\},
\]
i.e., the $J=3/2$ states precede the $J=1/2$ states, with $M$ increasing from $-J$ to $J$ within each $J$ block.

\begin{table}[t]
    \centering
    \renewcommand{\arraystretch}{1.35}
    \setlength{\tabcolsep}{14pt}
\begin{tabular}{c| cc}
    \toprule
    $K$
    & $O_{KQ}\!\left(\frac{3}{2},\frac{1}{2}\right)$
    & $\Omega_{KQ}\!\left(\frac{3}{2},\frac{1}{2}\right)$ \\  [0.4em]
    \midrule
    \hline
    \multirow{2}{*}{1}
    & TR odd & TR even \\
    & \textit{magnetic dipoles} & \textit{electric dipoles} \\ [0.6em]
    \midrule
    \multirow{2}{*}{2}
    & TR even & TR odd \\
    & \textit{electric quadrupoles} & \textit{magnetic quadrupoles} \\ 
    \bottomrule
\end{tabular}
    \caption{Properties under time reversal (TR) of the double-$J$ operators
    $O_{KQ}(JJ')$ and $\Omega_{KQ}(JJ')$, defined in
    Eqs.~\ref{eq:opON_pQ}--\ref{eq:opOmegaN_0Q}, for
    $J=3/2$ and $J'=1/2$.}
    \label{tab:TR}. 
\end{table}

\clearpage
\section{First principles methods}

In the following, we describe our first principles calculations. In A. we detail the electronic structure calculation of paramagnetic \bmro, in B. the calculation of Intersite Exchange Interactions (IEI) including generalized multipolar operators. In C. we detail the calculation of the Jahn-Teller coupling constants for both single-$J$ and double-$J$ operators and in D. the calculation of the classical electrostatic quadrupole-quadrupole interactions (cQQ). The intersite elastic couplings (IEC) have been taken from a previous study and their calculation is detailed in the supplemental material of Ref.~\cite{Fioremosca2024b} Sec. II. 

\subsection{Correlated electronic structure calculations}\label{ssec:dft+hi}

We compute the paramagnetic electronic structure of \bmro\ using fully charge-self-consistent density functional theory combined with dynamical mean-field theory (DFT+DMFT)~\cite{Georges1996,Anisimov1997_1,Lichtenstein_LDApp,Aichhorn2016}, implemented on top of the full-potential linearized augmented plane-wave (LAPW) code \textsc{Wien2k}~\cite{Wien2k}. Spin--orbit coupling is included in the DFT step through the standard second-variational method, and the DMFT impurity problem is solved within the quasi-atomic Hubbard-I (HI) approximation~\cite{hubbard_1}.
All calculations in the main text use the experimental cubic structure of \bmro, with lattice parameter $a = 8.0802$~\AA~\cite{Hirai2019_JPSJ}. We employ the local density approximation (LDA) for the exchange-correlation potential, 400 $\mathbf{k}$-points in the full Brillouin zone, and a basis cutoff $R_{\mathrm{MT}}K_{\mathrm{max}} = 7$.
Wannier orbitals for the Re $t_{2g}$ states are constructed from the Kohn--Sham bands of $t_{2g}$ character in the energy window $[-1.36, 2.04]$~eV around the Kohn--Sham Fermi level. The on-site Coulomb interaction within the $t_{2g}$ shell is described by the Kanamori Hamiltonian with $U_{\mathrm{K}} = 3.77$~eV and $J_{\mathrm{H}}^{\mathrm{K}} = 0.39$~eV. These values correspond to the full $d$-shell Slater parameters $F^0 = U = 3.2$~eV and $J_{\mathrm{H}} = 0.5$~eV, as used in our previous studies of $d^1$ double perovskites~\cite{Fioremosca2025, Fioremosca2024b}. The double-counting correction is treated in the fully localized limit with the nominal Re $5d$ occupancy $N_d = 1$.

\subsection{Intersite exchange interactions}

The IEI Hamiltonian, Eq.~(2) of the main text, is reported here for completeness. The IEI between a pair of Re ions on bond $ij$ takes the general form
\beq
\label{Eq:iei}
H_{ij}^{\rm IEI} = \sum_{\mu\mu'}\sum_{\nu\nu'} V_{\mu\mu'}^{\nu\nu'}(ij)\, \mathcal{O}_{\mu}^{\nu}(i)\, \mathcal{O}_{\mu'}^{\nu'}(j),
\eeq
where $V_{\mu\mu'}^{\nu\nu'}(ij)$ are the coupling constants and $\mathcal{O}^{\nu}_{\mu}$ are multipolar operators in compact notation; these operators are fully defined in Suppl.~Sec.~\ref{sec:ops}. The superscript $\nu\equiv\{JJ't\}$ labels the pair of $J$ manifolds and the operator type $t$, either normal ($t=n$) or anomalous ($t=a$). The subscript $\mu\equiv\{K,Q\}$ labels the multipolar rank and its projection. For $\nu=\{J,J,n\}$, the operators reduce to the conventional single-$J$ multipoles $O_{KQ}(J)$~\cite{Santini2009}.

The operators are defined in the basis $|JJ',M\rangle$, with $J=3/2$, $J'=1/2$, and $M=-J,\dots,J;\,-J',\dots,J'$. Within each manifold, the phases of the basis states are chosen such that $\langle JJ',M+1|J_+|JJ',M\rangle$ is real and positive. The relative phase between the two manifolds is fixed by requiring the matrix element of $S_+$ between $|\tfrac32,-\tfrac32\rangle = |m_l{=}-1\rangle\otimes|{\downarrow}\rangle$ and $|\tfrac12,-\tfrac12\rangle$ to equal its value for states constructed from $|l_{\rm eff}=1,m_l\rangle\otimes|s,\sigma\rangle$ with standard Clebsch--Gordan coefficients.

We evaluate the IEI from a converged paramagnetic electronic structure of \bmro using the force-theorem Hubbard-I (FT-HI) approach~\cite{Pourovskii2016}. In this method, small symmetry-breaking fluctuations of the density matrix within the full SOC manifold (\jeff=3/2 $\oplus$ \jeff=1/2) are applied simultaneously on two neighboring Re sites $i$ and $j$, and the couplings $V_{\mu\mu'}^{\nu\nu'}(ij)$ are obtained from the response of the DFT+DMFT grand potential to these two-site fluctuations. For double-$J$ operators, this response retains its original form,
\begin{equation}
    \frac{\delta^2 \Omega}{\delta \rho^{M_1 M_2}_{\vR}\, \delta \rho^{M_3 M_4}_{\vR+\Delta\vR}} \equiv \langle M_1 M_3 | V(\Delta\vR) | M_2 M_4 \rangle = \mathrm{Tr}\left[ G_{\Delta\vR}\, \frac{\delta \Sigma^{at}_{\vR+\Delta\vR}}{\delta\rho^{M_3 M_4}_{\vR+\Delta\vR}}\, G_{-\Delta\vR}\, \frac{\delta \Sigma^{at}_{\vR}}{\delta\rho^{M_1 M_2}_{\vR}} \right],
\label{eq:fluctuation}
\end{equation}
with the projection quantum numbers now running over both manifolds. Here $\rho^{M_1 M_2}_{\vR}$ is the corresponding element of the double-$J$ density matrix on site $\vR$, $G_{\Delta\vR}$ is the intersite Green's function, and $\delta\Sigma^{at}_{\vR}/\delta\rho^{M_1 M_2}_{\vR}$ is the derivative of the atomic self-energy with respect to the fluctuation $\rho^{M_1 M_2}_{\vR}$.

Using the matrix elements $[\mathcal{O}_\mu^\nu]^{M_1 M_2}$ of the real spherical tensor operators in the basis $|JJ',M\rangle$, normalized such that $\mathrm{Tr}[\mathcal{O}^{\nu}_{\mu}\mathcal{O}^{\nu'\dagger}_{\mu'}]=\delta_{\mu\mu'}\delta_{\nu\nu'}$, Eq.~\eqref{eq:fluctuation} is projected onto the couplings of Eq.~\eqref{Eq:iei}:
\begin{equation}
    V_{\mu\mu'}^{\nu\nu'}(\Delta\vR) = \sum_{\substack{M_1 M_2 \\ M_3 M_4}} \langle M_1 M_3 | V(\Delta\vR) | M_2 M_4 \rangle\, [\mathcal{O}_\mu^\nu]^{M_1 M_2}\, [\mathcal{O}_{\mu'}^{\nu'}]^{M_3 M_4}.
\end{equation}

The application of FT-HI to \bmro with double-$J$ operators is otherwise analogous to its previous applications to \jeff=3/2 double perovskites~\cite{Fioremosca2021, Fioremosca2024b}. Further details are given in the Supplemental Material of Ref.~\cite{Fioremosca2021}, and the full derivation of the method in Ref.~\cite{Pourovskii2016}.

\subsection{Jahn-Teller coupling constants}\label{sec:JT_coup}

To determine the JT coupling constants $g_{\Gamma}^{\nu}$ in eq. 5 of the main text we used the same approach as in Ref.~\onlinecite{Fioremosca2024b}, namely, we extracted them from DFT+HI calculations of distorted structures realizing the corresponding local distortions. The methodology is described in Suppl.~Sec.~IC of Ref.~\onlinecite{Fioremosca2024b}. In fact, we employed the same structure, see Suppl. Fig.~1 in \onlinecite{Fioremosca2024b}, implementing the $X_2^+$ distortion mode to extract the relevant $E_g$ JT coupling $g_{E_g}^{\nu}$ with $\nu= \{\tfrac{3}{2}\}\equiv 3/2$ or $ \{\tfrac{3}{2},\tfrac{1}{2},n\}\equiv{\mathrm{mix}}$ for the single-$J=3/2$ and mixed-$J$ quadrupole (where in the RHS we introduced the shortcut notation used below). The local distortion mode amplitude in this case is $q_{x^2-y^2}$ and the corresponding 
contribution due to the JT term to the on-site level position matrix reads
\beq\label{eq:eg_JT}
\epsilon_{\mathrm{JT}}=-\left[g^{3/2}_{E_g} O_{x^2-y^2}(\tfrac{3}{2})+g^{\mathrm{mix}}_{E_g} O_{x^2-y^2}(\tfrac{3}{2}\tfrac{1}{2})\right] q_{x^2-y^2},
\eeq
where the quadrupole operators $O_{x^2-y^2}(\tfrac{3}{2})$ and $O_{x^2-y^2}(\tfrac{3}{2}\tfrac{1}{2})$ are given explicitely at the end of Suppl. Sec~\ref{sec:ops}.

From the converged DFT+HI data~\cite{Fioremosca2024b} we obtained the converged level position matrices $\epsilon$ as a function of the $q_{x^2-y^2}$ amplitude. We transformed $\epsilon$, initially given in the $t_{2g}\otimes$S one-electron basis, into the $(J=3/2)\oplus(J=1/2)$ basis and expanded the result into the full set of multipolar  operators $\mathcal{O}_{\mu}^{\nu}$ defined above. The only substantial contribution for $K \ge 1$ is due to the JT term (the SOC contributes only to the monopole terms). Subsequently, the JT couplings $g^{3/2}_{E_g}$ and $g^{\mathrm{mix}}_{E_g}$ are extracted from linear fits of $\epsilon\cdot O_{x^2-y^2}(\tfrac{3}{2})$ and $\epsilon\cdot O_{x^2-y^2}(\tfrac{3}{2}\tfrac{3}{2})$ vs $q_{x^2-y^2}$, respectively. The resulting JT constants are listed in Suppl.~Table~\ref{tab:JT}.

\begin{table}[htbp]
\caption{Calculated JT coupling constants in meV/pm$^2$.
}\label{tab:JT}
\renewcommand{\arraystretch}{1.5}
\centering\begin{ruledtabular}\begin{tabular}{l  c  c  c  c  c}
Pressure (GPa) & -10 & -5 & 0 & 5 & 10 \\
\hline
$g_{E_g}^{3/2}$ & 13.79 & 14.43 & 15.06 & 15.97 & 16.65 \\
$g_{E_g}^{\mathrm{mix}}$ & -19.27 & -20.19 & -21.11 & -22.38 & -23.35 \\
\end{tabular}\end{ruledtabular}
\end{table}

One can also derive the $g_{E_g}^{\mathrm{mix}}/g_{E_g}^{3/2}$ ratio analytically by transforming  the $E_g$ JT Hamiltonian from the $t_{2g}\otimes S$ basis to the $(J=3/2)\oplus(J=1/2)$ one. This results in  $g_{E_g}^{\mathrm{mix}}/g_{E_g}^{3/2}=-\sqrt{2}$, see Ref.~\onlinecite{Teruki2026} for a detailed derivation. The same ratio is found between our calculated  JT couplings $g_{E_g}^{\mathrm{mix}}$ and $g_{E_g}^{3/2}$ listed in Table~\ref{tab:JT}.

\subsection{Classical quadrupole--quadrupole interactions}

\subsubsection*{Trial Field Calculation}

The cQQ couplings are obtained from fully charge self-consistent DFT+HI calculations in which a static single-particle field is applied to a \textit{trial} site $i_0$, while the induced quadrupolar field is read off at a \textit{measurement} site $i$ that carries no external field. Since the Hubbard-I approximation neglects the hybridization function, and consequently kinetic intersite exchange, the field acting on the measurement site contains only the classical electrostatic contribution. The perturbation
\begin{equation}
\hat H_Q = \Delta_Q \sum_{\alpha\beta} \Big[O_{2Q}(\tfrac{3}{2})\Big]_{\alpha\beta}\,
c^\dagger_{i_0\alpha} c^{\phantom{\dagger}}_{i_0\beta},
\end{equation}
acts on the six local spin-orbitals $\alpha,\beta$ of the (\jeff=3/2)$\oplus( $\jeff=1/2)
manifold. 
The perturbation is introduced directly into the local one-electron Hamiltonian,
\begin{equation}
  H^{\rm loc}_{\alpha\beta}(i) \longrightarrow H^{\rm loc}_{\alpha\beta}(i)
  + \delta_{i,i_0}\,\Delta_Q\,\Big[O_{2Q}(\tfrac{3}{2})\Big]_{\alpha\beta},
  \label{eqS:Hloc}
\end{equation}
so that any quadrupolar field acquired by the measurement site is generated entirely by the interaction with the polarized trial site and its periodic images.

The local correlated space is restricted to the $t_{2g}$ manifold, and we employ a 4 f.u.\ supercell with lattice constant $a=8.0802$~\AA. The trial site is located at $\mathbf{R}_{i_0}=[0.5,0.5,0]$ and the measurement site at $\mathbf{R}_i=[0,0,0]$. The couplings are converged with respect to the $\mathbf{k}$-point sampling; the results reported use a $6\times6\times6$ mesh. We consider $\Delta_Q=0.1$, 1, 10 and 25~meV for $Q=z^2$ and $x^2$-$y^2$, together with a reference calculation with $\Delta_Q=0.01$~meV. To eliminate the spurious self-polarization of the perturbed site we average the Boltzmann weights over the four states of the $J_{\rm eff}=3/2$ manifold to avoid the nonphysical self-interaction contribution~\cite{brooks1997density}.

Two quantities are extracted from each self-consistent solution. First, the induced quadrupoles at the trial site. From the local density matrix
\beq
  n_{\alpha\beta} = \langle c^{\dagger}_{\beta} c_{\alpha}\rangle
  = \frac{1}{\beta}\sum_n G^{\rm loc}_{\alpha\beta}(i\omega_n)\, e^{i\omega_n 0^+}
  \label{eqS:n}
\eeq
we obtain by rotation in the (\jeff=3/2)$\oplus( $\jeff=1/2) basis both the local density matrix $\rho$ and the level positions $\epsilon (i)$.
Second, the mean quadrupolar fields at the measurement site, obtained by projecting its converged local Hamiltonian  onto the same operators,
\beq
  B^{3/2}_{2Q}(i) = \mathrm{Tr}\,\big[O_{2Q}(\tfrac32)\, \epsilon (i)\big],
  \label{eqS:B}
\eeq
as detailed in the End Matter.

\subsubsection*{Mean field at the measurement site}

As discussed in the main text, the cQQ coupling constant is the linear-response coefficient
\beq
  I^{3/2}_{QQ'} = \frac{1}{N_{\rm im}}\left.\frac{\partial B^{3/2}_{2Q}(i)}{\partial\langle O_{2Q'}(\tfrac{3}{2})(i_0)\rangle}\right|_{\Delta_Q\to0},
\label{eq:V}
\eeq
where the factor $1/N_{\rm im}$, with $N_{\rm im}=4$, removes the multiplicity arising from the periodic images of the trial site. To see this, consider the most general quadrupole--quadrupole interaction matrix, previously derived from symmetry arguments in SM of Ref.~\onlinecite{Fioremosca2024b}, between the Re site at the origin and the Re site at $\vR=[0.5,0.5,0]$, restricting ourselves for simplicity to interactions between quadrupole operators $O_{2Q}(J)$ within a single $J$ multiplet:
\beq
\begin{tikzpicture}[baseline={([yshift=-0.6ex]current bounding box.center)}]
  \matrix (m) [matrix of math nodes, nodes in empty cells,
               row sep=1pt, column sep=2pt,
               nodes={minimum height=1.5em, minimum width=2.1em, anchor=center,
                      inner sep=1pt},
               column 1/.style={nodes={minimum width=0pt, anchor=east}}]
  {
    z^2     &[5pt] V^{1}_{A_g}      &                & V^{\rm off}_{A_g} &         &         \\
    x^2-y^2 &                   & V^{q}_{B_{1g}} &                   &         &         \\
    xy      & V^{\rm off}_{A_g} &                & V^{2}_{A_g}       &         &         \\
    xz      &                   &                &                   & V^{q}_1 & V^{q}_2 \\
    yz      &                   &                &                   & V^{q}_2 & V^{q}_1 \\
  };
  \coordinate (TL) at ([xshift=-1pt]m-1-2.north west);
  \coordinate (BL) at ([xshift=-1pt]m-5-2.south west);
  \coordinate (TR) at ([xshift=1pt]m-1-6.north east);
  \coordinate (BR) at ([xshift=1pt]m-5-6.south east);
  \draw ($(TL)+(3pt,0)$) -- (TL) -- (BL) -- ($(BL)+(3pt,0)$);
  \draw ($(TR)+(-3pt,0)$) -- (TR) -- (BR) -- ($(BR)+(-3pt,0)$);
  \coordinate (L) at ([xshift=-4pt]m-2-1.west);
  \coordinate (r1) at (m-1-2.north); \coordinate (r2) at (m-2-2.south);
  \coordinate (r3) at (m-3-2.north); \coordinate (r5) at (m-5-2.south);
  \draw ($(L|-r1)+(3pt,-1pt)$) -- ++(-3pt,0) -- ($(L|-r2)+(0,1pt)$) -- ++(3pt,0);
  \node[left] at ($(L|-r1)!0.5!(L|-r2)$) {$E_g^{+}$};
  \draw ($(L|-r3)+(3pt,-1pt)$) -- ++(-3pt,0) -- ($(L|-r5)+(0,1pt)$) -- ++(3pt,0);
  \node[left] at ($(L|-r3)!0.5!(L|-r5)$) {$T_{2g}^{+}$};
\end{tikzpicture}
\label{eq:cQQmatrix}
\eeq
In the supercell, the measurement site also interacts with the images of the trial site at $\vR=[-0.5,0.5,0]$, $[-0.5,-0.5,0]$, and $[0.5,-0.5,0]$, which are related to $[0.5,0.5,0]$ by the mirror planes $x\to-x$ and $y\to-y$. Summing the four contributions gives
\beq
\begin{tikzpicture}[baseline={([yshift=-0.6ex]current bounding box.center)}]
  \matrix (m) [matrix of math nodes, nodes in empty cells,
               row sep=1pt, column sep=2pt,
               nodes={minimum height=1.5em, minimum width=2.1em, anchor=center,
                      inner sep=1pt},
               column 1/.style={nodes={minimum width=0pt, anchor=east}}]
  {
    z^2     &[5pt] 4V^{1}_{A_g}     &                  &                &           &           \\
    x^2-y^2 &                   & 4V^{q}_{B_{1g}}  &                &           &           \\
    xy      &                   &                  & 4V^{2}_{A_g}   &           &           \\
    xz      &                   &                  &                & 4V^{q}_1  &           \\
    yz      &                   &                  &                &           & 4V^{q}_1  \\
  };
  \coordinate (TL) at ([xshift=-1pt]m-1-2.north west);
  \coordinate (BL) at ([xshift=-1pt]m-5-2.south west);
  \coordinate (TR) at ([xshift=1pt]m-1-6.north east);
  \coordinate (BR) at ([xshift=1pt]m-5-6.south east);
  \draw ($(TL)+(3pt,0)$) -- (TL) -- (BL) -- ($(BL)+(3pt,0)$);
  \draw ($(TR)+(-3pt,0)$) -- (TR) -- (BR) -- ($(BR)+(-3pt,0)$);
  \coordinate (L) at ([xshift=-4pt]m-2-1.west);
  \coordinate (r1) at (m-1-2.north); \coordinate (r2) at (m-2-2.south);
  \coordinate (r3) at (m-3-2.north); \coordinate (r5) at (m-5-2.south);
  \draw ($(L|-r1)+(3pt,-1pt)$) -- ++(-3pt,0) -- ($(L|-r2)+(0,1pt)$) -- ++(3pt,0);
  \node[left] at ($(L|-r1)!0.5!(L|-r2)$) {$E_g^{+}$};
  \draw ($(L|-r3)+(3pt,-1pt)$) -- ++(-3pt,0) -- ($(L|-r5)+(0,1pt)$) -- ++(3pt,0);
  \node[left] at ($(L|-r3)!0.5!(L|-r5)$) {$T_{2g}^{+}$};
\end{tikzpicture}
\label{eq:cQQmatrix_sum}
\eeq
Under these mirror operations, the off-diagonal couplings $V^{\rm off}_{A_g}$ ($z^2$--$xy$) and $V^{q}_{2}$ ($xz$--$yz$) change sign and cancel, whereas the diagonal couplings are invariant and add up. The field at the measurement site therefore contains each diagonal coupling four times, and dividing by $N_{\rm im}=4$ in Eq.~\eqref{eq:V} recovers the coupling of a single bond. We note that, in this geometry, only the diagonal couplings are accessible from the trial-field calculation.

\subsubsection*{Classical quadrupole-quadrupole couplings in \Jeff$= 3/2 \oplus 1/2$}

The cQQ couplings obtained within the \jeff$=3/2$ subspace are extended to the full (\jeff=3/2)$\oplus( $\jeff=1/2) manifold as described in the End Matter of the main text. In brief, the electrostatic interaction acts on the quadrupolar charge densities of the entire $t_{2g}$ shell, so the \jeff$=3/2$ coupling is its projection onto that subspace. Expressing the $t_{2g}$ quadrupole operators in the double-$J$ basis of Sec.~S1, $O^{t_{2g}}_{2Q}=O_{2Q}(\tfrac32)-\sqrt{2}\,O_{2Q}(\tfrac32\tfrac12)$, gives the couplings in the full manifold as $I^{\nu\nu'}_{2Q2Q'} = c_\nu c_{\nu'}\, I^{3/2}_{QQ'}$, with $c_{3/2}=1$ and $c_{3/2\,1/2}=-\sqrt{2}$. The $t_{2g}$ quadrupoles have no component within the \jeff$=1/2$ doublet.

In the following, we report the cQQ couplings in the full (\jeff=3/2)$\oplus( $\jeff=1/2) SOC manifold.

\begin{table}[htb]
\setlength{\tabcolsep}{8pt}
\renewcommand{\arraystretch}{1.5}
\caption{cQQ couplings per bond (meV) between the generalized quadrupoles for the in-plane
bond $\mathbf{R}=[0.5,\,0.5,\,0]$, with the shorthand $\tfrac32\equiv(\tfrac32,\tfrac32,n)$
and $\tfrac32\tfrac12\equiv(\tfrac32,\tfrac12,n)$.}
\label{tabS:Iup}
\begin{tabular}{| c| ccc | }
\hline \hline
 & $I^{\frac32,\frac32}_{2Q,2Q}$ & $I^{\frac32,\frac32\frac12}_{2Q,2Q}$ &
 $I^{\frac32\frac12,\frac32\frac12}_{2Q,2Q}$ \\
\hline
$z^2$       & $1.89$  & $-2.67$ & $3.78$  \\
\hline
$x^2$-$y^2$ & $-0.86$ & $1.22$  & $-1.72$ \\
\hline \hline
\end{tabular}
\end{table}

\subsubsection*{Classical Quadrupole-quadrupole interactions in \jeff = 3/2 under pressure}

In the following we report the cQQ couplings $I^{\frac32,\frac32}_{2Q,2Q}$ as a function of pressure, used for the calculation of the pressure phase diagram in the End Matter. The mixed cQQ couplings can be readily derived by considering the relation of Eq. 13 in the End Matter.

\begin{table}[htbp]
\caption{Calculated cQQ coupling constants in meV.
}\label{tab:JT}
\renewcommand{\arraystretch}{1.5}
\centering\begin{ruledtabular}\begin{tabular}{l  c  c  c  c  c}
Pressure (GPa) & -10 & -5 & 0 & 5 & 10 \\
\hline
$z^2$ & 1.810 & 1.855 & 1.876 & 1.912 & 1.946 \\
$x^2-y^2$ & -0.902 & -0.881 & -0.860 & -0.841 & -0.822 \\
\end{tabular}\end{ruledtabular}
\end{table}

\section{Intersite exchange interactions}

In Suppl. Table~\ref{tab:bmro-exchange-110} we list the calculated IEI couplings for the representative NN lattice vector $\vR=$[0.5,0.5,0] for zero pressure as well  as for $P=\pm10$~GPa. The values within  the \jeff=3/2 manifold coincide, up to rounding errors, with those listed in the SM of Ref.~\onlinecite{Fioremosca2024b}. In addition, Suppl. Table~\ref{tab:bmro-exchange-110}  includes IEI between all double-$J$ multipoles and those coupling double-$J$ and \jeff=3/2 multipoles. Since the single-\jeff=1/2 degrees of freedom (monopole and dipoles) do not play any noticeable in  the ordered phases of BMRO, we skip the corresponding IEI for brevity. 

One may notice that the  calculated IEI agree with the multipole symmetry under time reversal discussed in Suppl.~Sec.~\ref{sec:ops}. In particular, the anomalous electrical dipoles $\Omega_{1Q}$ couple to both the single-$J$ and double-$J$ electrical quadrupoles $O_{2Q}$, while the anomalous magnetic quadrupoles $\Omega_{2Q}$ couple to the magnetic dipoles $O_{1Q}$ and octupoles $O_{3Q}$.

\ifdefined\BMROExchangeCaptionBox\else\newsavebox{\BMROExchangeCaptionBox}\fi
\ifdefined\BMROExchangePanelsBox\else\newsavebox{\BMROExchangePanelsBox}\fi
\ifdefined\BMROExchangeAvailableHeight\else\newlength{\BMROExchangeAvailableHeight}\fi
\begin{table*}[p]
\centering
\setlength{\parskip}{0pt}
\setlength{\parindent}{0pt}
\begin{lrbox}{\BMROExchangeCaptionBox}
\begin{minipage}{\textwidth}
\fontsize{8.5}{9.5}\selectfont

\caption{Calculated IEI for the lattice vector $\vR_{ij}$=[0.5,0.5,0], in meV. The IEI involving single-\jeff=1/2 degrees of freedom are not included. Since the IEI matrices are symmetric, only the lower triangle values are listed. Only the values above 0.01~meV are shown.  }\label{tab:bmro-exchange-110}
\end{minipage}
\end{lrbox}
\setlength{\BMROExchangeAvailableHeight}{\dimexpr\textheight-\ht\BMROExchangeCaptionBox-\dp\BMROExchangeCaptionBox-8pt\relax}
\begin{lrbox}{\BMROExchangePanelsBox}
\begin{minipage}{\textwidth}
\centering
\begingroup
\fontsize{8}{8.2}\selectfont
\setlength{\tabcolsep}{7.2pt}
\setlength{\heavyrulewidth}{0.64pt}
\setlength{\lightrulewidth}{0.40pt}
\setlength{\cmidrulewidth}{0.24pt}
\setlength{\cmidrulekern}{4pt}
\setlength{\abovetopsep}{0pt}
\setlength{\belowbottomsep}{0pt}
\setlength{\cmidrulesep}{2pt}
\setlength{\aboverulesep}{0.4pt}
\setlength{\belowrulesep}{0.7pt}
\renewcommand{\arraystretch}{1.0}
\begin{minipage}[t]{0.485\textwidth}
\vspace{0pt}
\centering
\begin{tabular}{@{}rrrrrrr@{}}
\toprule
\multicolumn{4}{c}{Multipoles} & \multicolumn{3}{c}{Pressure (GPa)}\\
\cmidrule(lr){1-4}\cmidrule(lr){5-7}
$K$ & $Q$ & $K^{\prime}$ & $Q^{\prime}$ & $10$ & $0$ & $-10$\\
\midrule
\multicolumn{7}{l}{$\mathcal{O}_{\mu}^{\nu}(i)=O_{KQ}(\tfrac32),\quad \mathcal{O}_{\mu'}^{\nu'}(j)=O_{K'Q'}(\tfrac32)$}\\
\cmidrule(lr){1-7}
\multicolumn{7}{l}{Monopole--Monopole}\\
$0$ & $0$ & $0$ & $0$ & $-0.61$ & $-0.56$ & $-0.40$ \\
\cmidrule(lr){1-7}
\multicolumn{7}{l}{Dipole--Dipole}\\
$1$ & $-1$ & $1$ & $-1$ & $-0.80$ & $-0.59$ & $-0.37$ \\
$1$ & $0$ & $1$ & $0$ & $0.86$ & $0.73$ & $0.45$ \\
$1$ & $1$ & $1$ & $-1$ & $0.87$ & $0.91$ & $0.77$ \\
$1$ & $1$ & $1$ & $1$ & $-0.80$ & $-0.59$ & $-0.37$ \\
\cmidrule(lr){1-7}
\multicolumn{7}{l}{Quadrupole--Monopole}\\
$2$ & $-2$ & $0$ & $0$ & $-0.41$ & $-0.44$ & $-0.37$ \\
$2$ & $0$ & $0$ & $0$ & $1.12$ & $0.98$ & $0.68$ \\
\cmidrule(lr){1-7}
\multicolumn{7}{l}{Quadrupole--Quadrupole}\\
$2$ & $-2$ & $2$ & $-2$ & $-1.82$ & $-1.60$ & $-1.07$ \\
$2$ & $-1$ & $2$ & $-1$ & $-2.38$ & $-2.15$ & $-1.53$ \\
$2$ & $0$ & $2$ & $-2$ & $1.05$ & $1.08$ & $0.90$ \\
$2$ & $0$ & $2$ & $0$ & $2.72$ & $2.45$ & $1.73$ \\
$2$ & $1$ & $2$ & $-1$ & $-0.10$ & $-0.10$ & $-0.08$ \\
$2$ & $1$ & $2$ & $1$ & $-2.38$ & $-2.15$ & $-1.53$ \\
$2$ & $2$ & $2$ & $2$ & $-2.08$ & $-1.90$ & $-1.36$ \\
\cmidrule(lr){1-7}
\multicolumn{7}{l}{Octupole--Dipole}\\
$3$ & $-3$ & $1$ & $-1$ & $0.06$ & $0.08$ & $0.07$ \\
$3$ & $-3$ & $1$ & $1$ & $-0.31$ & $-0.31$ & $-0.25$ \\
$3$ & $-2$ & $1$ & $0$ & $0.59$ & $0.61$ & $0.50$ \\
$3$ & $-1$ & $1$ & $-1$ & $-1.92$ & $-1.80$ & $-1.28$ \\
$3$ & $-1$ & $1$ & $1$ & $-0.21$ & $-0.21$ & $-0.18$ \\
$3$ & $0$ & $1$ & $0$ & $-0.08$ & $-0.12$ & $-0.09$ \\
$3$ & $1$ & $1$ & $-1$ & $-0.21$ & $-0.21$ & $-0.18$ \\
$3$ & $1$ & $1$ & $1$ & $-1.92$ & $-1.80$ & $-1.28$ \\
$3$ & $3$ & $1$ & $-1$ & $0.31$ & $0.31$ & $0.25$ \\
$3$ & $3$ & $1$ & $1$ & $-0.06$ & $-0.08$ & $-0.07$ \\
\cmidrule(lr){1-7}
\multicolumn{7}{l}{Octupole--Octupole}\\
$3$ & $-3$ & $3$ & $-3$ & $1.03$ & $0.88$ & $0.57$ \\
$3$ & $-2$ & $3$ & $-2$ & $-1.62$ & $-1.43$ & $-0.96$ \\
$3$ & $-1$ & $3$ & $-3$ & $-0.07$ & $-0.09$ & $-0.09$ \\
$3$ & $-1$ & $3$ & $-1$ & $1.04$ & $1.05$ & $0.81$ \\
$3$ & $0$ & $3$ & $-2$ & $-0.55$ & $-0.60$ & $-0.52$ \\
$3$ & $0$ & $3$ & $0$ & $2.69$ & $2.58$ & $1.92$ \\
$3$ & $1$ & $3$ & $-3$ & $-0.28$ & $-0.28$ & $-0.22$ \\
$3$ & $1$ & $3$ & $-1$ & $-0.72$ & $-0.76$ & $-0.65$ \\
$3$ & $1$ & $3$ & $1$ & $1.04$ & $1.05$ & $0.81$ \\
$3$ & $2$ & $3$ & $2$ & $-1.81$ & $-1.66$ & $-1.18$ \\
$3$ & $3$ & $3$ & $-1$ & $0.28$ & $0.28$ & $0.22$ \\
$3$ & $3$ & $3$ & $1$ & $0.07$ & $0.09$ & $0.09$ \\
$3$ & $3$ & $3$ & $3$ & $1.03$ & $0.88$ & $0.57$ \\
\midrule
\multicolumn{7}{l}{$\mathcal{O}_{\mu}^{\nu}(i)=O_{KQ}(\tfrac{3}{2}\tfrac{1}{2}),\quad \mathcal{O}_{\mu'}^{\nu'}(j)=O_{K'Q'}(\tfrac32)$}\\
\cmidrule(lr){1-7}
\multicolumn{7}{l}{Dipole--Dipole}\\
$1$ & $-1$ & $1$ & $-1$ & $-2.38$ & $-2.19$ & $-1.54$ \\
$1$ & $-1$ & $1$ & $1$ & $0.50$ & $0.52$ & $0.43$ \\
$1$ & $0$ & $1$ & $0$ & $0.65$ & $0.70$ & $0.54$ \\
$1$ & $1$ & $1$ & $-1$ & $-0.50$ & $-0.52$ & $-0.43$ \\
$1$ & $1$ & $1$ & $1$ & $2.38$ & $2.19$ & $1.54$ \\
\cmidrule(lr){1-7}
\multicolumn{7}{l}{Dipole--Octupole}\\
$1$ & $-1$ & $3$ & $-3$ & $0.05$ & $0.06$ & $0.06$ \\
$1$ & $-1$ & $3$ & $-1$ & $0.24$ & $0.30$ & $0.25$ \\
$1$ & $-1$ & $3$ & $1$ & $-0.61$ & $-0.63$ & $-0.53$ \\
$1$ & $-1$ & $3$ & $3$ & $-0.21$ & $-0.21$ & $-0.17$ \\
$1$ & $0$ & $3$ & $-2$ & $0.11$ & $0.11$ & $0.09$ \\
$1$ & $0$ & $3$ & $0$ & $-3.86$ & $-3.62$ & $-2.64$ \\
$1$ & $1$ & $3$ & $-3$ & $-0.21$ & $-0.21$ & $-0.17$ \\
$1$ & $1$ & $3$ & $-1$ & $0.61$ & $0.63$ & $0.53$ \\
$1$ & $1$ & $3$ & $1$ & $-0.24$ & $-0.30$ & $-0.25$ \\
$1$ & $1$ & $3$ & $3$ & $0.05$ & $0.06$ & $0.06$ \\
\cmidrule(lr){1-7}
\multicolumn{7}{l}{Quadrupole--Monopole}\\
$2$ & $-2$ & $0$ & $0$ & $-0.43$ & $-0.44$ & $-0.37$ \\
$2$ & $0$ & $0$ & $0$ & $0.20$ & $0.20$ & $0.12$ \\
\cmidrule(lr){1-7}
\multicolumn{7}{l}{Quadrupole--Quadrupole}\\
$2$ & $-2$ & $2$ & $-2$ & $-2.47$ & $-2.18$ & $-1.52$ \\
$2$ & $-2$ & $2$ & $0$ & $0.41$ & $0.43$ & $0.38$ \\
$2$ & $-1$ & $2$ & $-1$ & $-1.46$ & $-1.26$ & $-0.86$ \\
$2$ & $-1$ & $2$ & $1$ & $-1.14$ & $-1.17$ & $-0.98$ \\
$2$ & $0$ & $2$ & $-2$ & $-0.35$ & $-0.34$ & $-0.29$ \\
$2$ & $0$ & $2$ & $0$ & $-4.77$ & $-4.37$ & $-3.22$ \\
$2$ & $1$ & $2$ & $-1$ & $1.14$ & $1.17$ & $0.98$ \\
$2$ & $1$ & $2$ & $1$ & $1.46$ & $1.26$ & $0.86$ \\
$2$ & $2$ & $2$ & $2$ & $-2.27$ & $-1.94$ & $-1.30$ \\
\bottomrule
\end{tabular}
\end{minipage}%
\hfill
\begin{minipage}[t]{0.485\textwidth}
\vspace{0pt}
\centering
\begin{tabular}{@{}rrrrrrr@{}}
\toprule
\multicolumn{4}{c}{Multipoles} & \multicolumn{3}{c}{Pressure (GPa)}\\
\cmidrule(lr){1-4}\cmidrule(lr){5-7}
$K$ & $Q$ & $K^{\prime}$ & $Q^{\prime}$ & $10$ & $0$ & $-10$\\
\midrule
\multicolumn{7}{l}{$\mathcal{O}_{\mu}^{\nu}(i)=\Omega_{KQ}(\tfrac{3}{2}\tfrac{1}{2}),\quad \mathcal{O}_{\mu'}^{\nu'}(j)=O_{K'Q'}(\tfrac32)$}\\
\cmidrule(lr){1-7}
\multicolumn{7}{l}{Dipole--Quadrupole}\\
$1$ & $-1$ & $2$ & $-1$ & $0.57$ & $0.60$ & $0.50$ \\
$1$ & $-1$ & $2$ & $1$ & $-1.95$ & $-1.67$ & $-1.12$ \\
$1$ & $0$ & $2$ & $2$ & $1.15$ & $1.19$ & $1.00$ \\
$1$ & $1$ & $2$ & $-1$ & $1.95$ & $1.67$ & $1.12$ \\
$1$ & $1$ & $2$ & $1$ & $-0.57$ & $-0.60$ & $-0.50$ \\
\cmidrule(lr){1-7}
\multicolumn{7}{l}{Quadrupole--Dipole}\\
$2$ & $-1$ & $1$ & $-1$ & $0.41$ & $0.41$ & $0.35$ \\
$2$ & $-1$ & $1$ & $1$ & $1.25$ & $1.24$ & $0.93$ \\
$2$ & $1$ & $1$ & $-1$ & $-1.25$ & $-1.24$ & $-0.93$ \\
$2$ & $1$ & $1$ & $1$ & $-0.41$ & $-0.41$ & $-0.35$ \\
$2$ & $2$ & $1$ & $0$ & $0.03$ & $0.02$ & $0.00$ \\
\cmidrule(lr){1-7}
\multicolumn{7}{l}{Quadrupole--Octupole}\\
$2$ & $-2$ & $3$ & $2$ & $1.91$ & $1.65$ & $1.11$ \\
$2$ & $-1$ & $3$ & $-3$ & $0.16$ & $0.16$ & $0.13$ \\
$2$ & $-1$ & $3$ & $-1$ & $-0.20$ & $-0.22$ & $-0.18$ \\
$2$ & $-1$ & $3$ & $1$ & $-3.71$ & $-3.43$ & $-2.48$ \\
$2$ & $-1$ & $3$ & $3$ & $0.08$ & $0.10$ & $0.10$ \\
$2$ & $0$ & $3$ & $2$ & $-0.97$ & $-1.02$ & $-0.85$ \\
$2$ & $1$ & $3$ & $-3$ & $0.08$ & $0.10$ & $0.10$ \\
$2$ & $1$ & $3$ & $-1$ & $3.71$ & $3.43$ & $2.48$ \\
$2$ & $1$ & $3$ & $1$ & $0.20$ & $0.22$ & $0.18$ \\
$2$ & $1$ & $3$ & $3$ & $0.16$ & $0.16$ & $0.13$ \\
$2$ & $2$ & $3$ & $-2$ & $2.04$ & $1.82$ & $1.26$ \\
$2$ & $2$ & $3$ & $0$ & $0.90$ & $0.93$ & $0.77$ \\
\midrule
\multicolumn{7}{l}{$\mathcal{O}_{\mu}^{\nu}(i)=O_{KQ}(\tfrac{3}{2}\tfrac{1}{2}),\quad \mathcal{O}_{\mu'}^{\nu'}(j)=O_{K'Q'}(\tfrac{3}{2}\tfrac{1}{2})$}\\
\cmidrule(lr){1-7}
\multicolumn{7}{l}{Dipole--Dipole}\\
$1$ & $-1$ & $1$ & $-1$ & $-0.35$ & $-0.25$ & $-0.17$ \\
$1$ & $0$ & $1$ & $0$ & $3.36$ & $3.13$ & $2.26$ \\
$1$ & $1$ & $1$ & $-1$ & $-0.31$ & $-0.31$ & $-0.25$ \\
$1$ & $1$ & $1$ & $1$ & $-0.35$ & $-0.25$ & $-0.17$ \\
\cmidrule(lr){1-7}
\multicolumn{7}{l}{Quadrupole--Quadrupole}\\
$2$ & $-2$ & $2$ & $-2$ & $-0.54$ & $-0.46$ & $-0.31$ \\
$2$ & $-1$ & $2$ & $-1$ & $-3.01$ & $-2.60$ & $-1.85$ \\
$2$ & $0$ & $2$ & $-2$ & $0.30$ & $0.29$ & $0.23$ \\
$2$ & $0$ & $2$ & $0$ & $6.98$ & $6.25$ & $4.57$ \\
$2$ & $1$ & $2$ & $-1$ & $0.87$ & $0.89$ & $0.74$ \\
$2$ & $1$ & $2$ & $1$ & $-3.01$ & $-2.60$ & $-1.85$ \\
$2$ & $2$ & $2$ & $2$ & $-0.70$ & $-0.64$ & $-0.48$ \\
\midrule
\multicolumn{7}{l}{$\mathcal{O}_{\mu}^{\nu}(i)=\Omega_{KQ}(\tfrac{3}{2}\tfrac{1}{2}),\quad \mathcal{O}_{\mu'}^{\nu'}(j)=O_{K'Q'}(\tfrac{3}{2}\tfrac{1}{2})$}\\
\cmidrule(lr){1-7}
\multicolumn{7}{l}{Dipole--Quadrupole}\\
$1$ & $-1$ & $2$ & $-1$ & $-0.36$ & $-0.37$ & $-0.31$ \\
$1$ & $-1$ & $2$ & $1$ & $-0.93$ & $-0.78$ & $-0.55$ \\
$1$ & $0$ & $2$ & $2$ & $0.72$ & $0.74$ & $0.62$ \\
$1$ & $1$ & $2$ & $-1$ & $-0.93$ & $-0.78$ & $-0.55$ \\
$1$ & $1$ & $2$ & $1$ & $-0.36$ & $-0.37$ & $-0.31$ \\
\cmidrule(lr){1-7}
\multicolumn{7}{l}{Quadrupole--Dipole}\\
$2$ & $-1$ & $1$ & $-1$ & $-0.04$ & $-0.04$ & $-0.03$ \\
$2$ & $-1$ & $1$ & $1$ & $1.81$ & $1.66$ & $1.19$ \\
$2$ & $1$ & $1$ & $-1$ & $1.81$ & $1.66$ & $1.19$ \\
$2$ & $1$ & $1$ & $1$ & $-0.04$ & $-0.04$ & $-0.03$ \\
$2$ & $2$ & $1$ & $0$ & $0.22$ & $0.22$ & $0.17$ \\
\midrule
\multicolumn{7}{l}{$\mathcal{O}_{\mu}^{\nu}(i)=\Omega_{KQ}(\tfrac{3}{2}\tfrac{1}{2}),\quad \mathcal{O}_{\mu'}^{\nu'}(j)=\Omega_{K'Q'}(\tfrac{3}{2}\tfrac{1}{2})$}\\
\cmidrule(lr){1-7}
\multicolumn{7}{l}{Dipole--Dipole}\\
$1$ & $-1$ & $1$ & $-1$ & $-1.40$ & $-1.23$ & $-0.88$ \\
$1$ & $0$ & $1$ & $0$ & $-3.10$ & $-2.67$ & $-1.89$ \\
$1$ & $1$ & $1$ & $-1$ & $-0.63$ & $-0.65$ & $-0.55$ \\
$1$ & $1$ & $1$ & $1$ & $-1.40$ & $-1.23$ & $-0.88$ \\
\cmidrule(lr){1-7}
\multicolumn{7}{l}{Quadrupole--Quadrupole}\\
$2$ & $-2$ & $2$ & $-2$ & $-0.71$ & $-0.65$ & $-0.47$ \\
$2$ & $-1$ & $2$ & $-1$ & $3.45$ & $3.14$ & $2.30$ \\
$2$ & $0$ & $2$ & $-2$ & $0.63$ & $0.65$ & $0.55$ \\
$2$ & $0$ & $2$ & $0$ & $-2.72$ & $-2.35$ & $-1.66$ \\
$2$ & $1$ & $2$ & $-1$ & $0.21$ & $0.21$ & $0.16$ \\
$2$ & $1$ & $2$ & $1$ & $3.45$ & $3.14$ & $2.30$ \\
$2$ & $2$ & $2$ & $2$ & $-0.61$ & $-0.53$ & $-0.36$ \\
\bottomrule
\end{tabular}
\end{minipage}%
\endgroup
\end{minipage}
\end{lrbox}
\usebox{\BMROExchangeCaptionBox}\par\nointerlineskip
\ifdim\dimexpr\ht\BMROExchangePanelsBox+\dp\BMROExchangePanelsBox\relax>\BMROExchangeAvailableHeight
  \resizebox*{!}{\BMROExchangeAvailableHeight}{\usebox{\BMROExchangePanelsBox}}%
\else
  \usebox{\BMROExchangePanelsBox}%
\fi
\end{table*}

\section{Mean-field calculations}

To solve the effective low-energy Hamiltonian, eq.~1 of the main text, we employ a generalized single-site mean-field approach. All intersite terms -- IEI, cQQ, and IEC -- are mean-field decoupled resulting in a single-site JT  problem in electronic (exchange+electrostatic)
and displacement mean field. This single-site problem is solved by introducing a vibronic basis $\{\phi\}$~\cite{Bersuker1989}, whose states are products $|M_i\rangle|n_{z^2},n_{x^2-y^2}\rangle$ of a $(J=3/2)\oplus(J=1/2)$ total-angular-momentum eigenstate and an $E_g$ local phonon wavefunction. The latter is specified by the occupation numbers $n_{z^2}$ and $n_{x^2-y^2}$ of the corresponding $E_g$ modes. The vibronic basis is cut at the maximum number $N_{\mathrm{ph}}^{\mathrm{MAX}}=\mathrm{MAX}\left(n_{z^2}+n_{x^2-y^2}\right)$ of local $E_g$ pnonons.  

The overall approach is rather standard~\cite{Englman70,Gehring_1975,Polinger2009,Iwahara2024} and is described in detail in Suppl.~Sec.~VII of Ref.~\onlinecite{Fioremosca2024b}. Here, we only highlight  the differences from the implementation of Ref.~\onlinecite{Fioremosca2024b}. In particular, we now include JT coupling to both the single-$J$ quadrupoles $O_{\mu}(\tfrac{3}{2})$ and the mixed quadrupoles $O_{\mu}(\tfrac{3}{2}\tfrac{1}{2})$, where $\mu=z^2$ and $x^2-y^2$, as well as the spin-orbit term $H^{\mathrm{SO}}$. Moreover, the electronic mean field now includes mixed multipoles as well as both exchange and classical electrostatic contributions. The single-site JT problem (cf. eq.~11 in Supplementary of Ref.~\onlinecite{Fioremosca2024b}) then reads:
\begin{equation}\label{eq:H1s}
H_{1s} = \sum_{\mu \in E_g}\left[ \hbar \omega_{E_g}(\hat{n}_{E_g \mu}+1/2) +  \left(\Phi_{E_g \mu}^{\mathrm{MF}}-g_{\Gm}^{3/2}O_{\mu}(\tfrac{3}{2})-g_{\Gm}^{\mathrm{mix}}O_{\mu}(\tfrac{3}{2}\tfrac{1}{2})\right)A_{E_g}(a^{\dagger}_{\mu}+a_{\mu})\right]+\sum_{\mu\nu}V_{\mu\nu}^{\mathrm{MF}}\mathcal{O}_{\mu}^{\nu}+H^{\mathrm{SO}},
\end{equation}
where we use the  JT coupling notation introduced in Sec.~\ref{sec:JT_coup}, the  sum in the first RHS term runs over $E_g$ modes and involves both $J=3/2$ and mixed quadrupoles for a given $\mu=z^2$, $x^2-y^2$. In the last term the sum runs over all $J=3/2$ and mixed multipole operators $\mathcal{O}_{\mu}^{\nu}$ and the corresponding mean field $V^{MF}_{\mu\nu}$ includes, for $J=3/2$ quadrupoles, also the cQQ contribution. One subsequently evaluates the matrix elements of all electronic multipolar operators  and phonon creation and annihilation operators in the vibronic basis $\{\phi\}$ as detailed in Ref.~\onlinecite{Fioremosca2024b}. 

In practice, the mean-field calculations were carried out using the McPhase package~\cite{Rotter2004}, with the JT problem implemented as a single-ion module. To simplify the implementation, we first diagonalize the single-site JT problem without the elastic and electronic mean-field terms, using the vibronic basis $\{\phi\}$ with a chosen cutoff $N_{\mathrm{ph}}^{\mathrm{MAX}}$. This yields the JT eigenvalues $E^{\mathrm{JT}}_i$ and eigenstates $\psi_i$. We then evaluate the displacement-operator matrix elements $D_{ij}(\mu)=\langle \psi_i |A_{E_g}(a^{\dagger}_{\mu}+a_{\mu})|\psi_j\rangle$ and the multipolar matrix elements $\widetilde{\mathcal{O}}_{ij}(\mu\nu)=\langle \psi_i |\mathcal{O}_{\mu}^{\nu}|\psi_j\rangle$. The single-site JT problem (\ref{eq:H1s}) then takes a particularly simple form in this JT basis:
$$
[H_{1s}]_{ij}= E_i^{\mathrm{JT}}\delta_{ij}+\sum_{\mu\in E_g}\Phi_{E_g \mu}^{\mathrm{MF}}D_{ij}(\mu)+\sum_{\mu\nu}V_{\mu\nu}^{\mathrm{MF}}\widetilde{\mathcal{O}}_{ij}(\mu\nu),
$$
and it is implemented in this form in the McPhase module. 

As
initial guesses of the mean-field procedure we employ all 1$\vk$ structures realizable within the BMRO single conventional
unit cell that contains 4 magnetic Re sites, with ten  random configuration flips  implemented by "McPhase" employed for each structure. 

\clearpage
\subsubsection*{Evolution of primary and secondary order parameters in \bmro as a function of $n_{ph}$}

In the following we show the convergence of the order parameters (primary: $\langle x^2 - y^2 \rangle \left( \tfrac32 \right) $, $\langle x^2 - y^2 \rangle \left( \tfrac32 \tfrac12 \right) $ and secondary: $\langle z^2 \rangle \left( \tfrac32 \right) $, $\langle z^2 \rangle \left( \tfrac32 \tfrac12 \right) $) as a function of the number of phonons. The primary order parameter $\langle x^2 - y^2 \rangle \left( \tfrac32 \right) $ is already converged at $n_{ph} = 2$, but the secondary ones are not yet converged, following the evolution of the other primary (mixed) order parameter $\langle x^2 - y^2 \rangle \left( \tfrac32 \tfrac12 \right) $.

\begin{figure}[!h]
  \begin{center}
  \includegraphics[width=0.5\textwidth, clip=true]{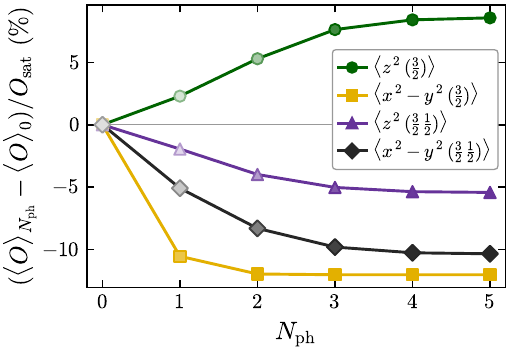}
    \end{center}
\caption{Evolution of the mean field order parameters normalized by their saturated values as a function of the cutoff $N_{ph}$ number of phonon in the vibronic basis. }
\label{fig:suppl1} 
\end{figure}


\bibliography{bibliography}